\documentclass[journal=jpclcd]{achemso}
\usepackage{amsmath}
\usepackage{amsfonts}
\usepackage{amssymb}
\usepackage{graphicx}
\usepackage{color}
\usepackage{lscape}

\usepackage{bm}
\usepackage{lmodern}
\newcommand{\mol}[1]{\ensuremath{_{\text{#1}}}}
\newcommand{\mr}[1]{\ensuremath{\mathrm{#1}}}
\renewcommand{\vec}[1]{\ensuremath{\bm{#1}}}

\title{Enhancing the efficiency of density functionals with a novel iso-orbital indicator.}

\author{James W. Furness}
\email{jfurness@tulane.edu}
\author{Jianwei Sun}
\email{jsun@tulane.edu}
\affiliation{Department of Physics and Engineering Physics, Tulane University, New Orleans, LA 70118, USA}

\date{\today}

\begin{document}

\begin{abstract}
The accuracy and efficiency of a density functional is dependent on the basic ingredients it uses, and how the ingredients are built into the functional as a whole. An iso-orbital indicator based on the electron density, its gradients, and the kinetic energy density, has proven an essential dimensionless variable that allows density functionals to recognise and correctly treat various types of chemical bonding, both strong and weak. Density functionals constructed around the iso-orbital indicator usually require dense real-space grids for numerical implementation that deteriorate computational efficiency, with poor grid convergence compromising the improved accuracy. Here, a novel iso-orbital indicator is proposed based on the same ingredients that retains the capability to identify the same chemical bonds while significantly relieving the requirement of dense grids. Furthermore, the novel iso-orbital indicator gives an improved recognition for tail regions of electron densities and is constraint-free for the exchange-correlation potential. The novel iso-orbital indicator is therefore expected to be the prime choice for further density functional development.

\begin{tocentry}
\includegraphics{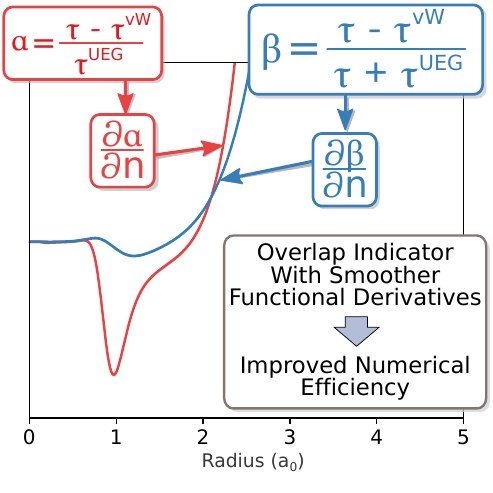}
\end{tocentry}
\end{abstract}

\maketitle

\section{Introduction}

Density functional theory (DFT) is, in principle, exact for the ground state energy and electron density of a system of electrons under a scalar external potential, conventionally solved through a set of Kohn--Sham (KS) auxiliary single-particle Schr\"odinger-like equations\cite{Hohenberg1964, Kohn1965}. In practice however, the exchange-correlation energy, an essential but usually small portion of the total energy, must be approximated as a functional of the electron density. The computational efficiency of this scheme and the accuracy of modern exchange-correlation approximations has resulted in DFT becoming one of the most widely used electronic structure theories and arguably the only practical method for high-throughput discovery of novel materials currently available.

Exchange-correlation approximations can be broadly categorised by the ingredients used into 5 levels of increasing non-locality\cite{Perdew2001}. The accuracy of an approximation usually increases when more ingredients are included through increased flexibility though this enhancement is often accompanied by a deterioration of efficiency, especially when non-local information is included. It is therefore critical to understand the ingredients of common density functional approximations and how they can be utilised to increase the accuracy of a functional whilst maximising the possible computational efficiency. As functionals of higher levels are usually developed based on functionals of lower levels, knowledge obtained for the ingredients and their combinations in lower-level functionals can be transferred to the development of more complex functionals at higher levels.

The lowest three levels of efficient semi-local functionals include the local spin density approximation (LSDA) \cite{Kohn1965, Ma1968, vonBarth1972, Perdew1981, Perdew1992a, Sun2010}, the generalised gradient approximation (GGA) \cite{Ma1968, Becke1988, Perdew1992, Perdew1996,Keal2004,Armiento2005,Perdew2008a, Constantin2011, Vela2012}, and meta-GGA \cite{Becke1989,VanVoorhis1998, Becke1998,Perdew1999,Tao2003,Becke2003,Zhao2006,Perdew2009,Peverati2012,Yu2016a, Sun2012, DelCampo2012, Mardirossian2015, Sun2015}. LSDA uses only the electron density and recovers the uniform electron gas (UEG) limit. GGAs add the electron density gradient from which two standard dimensionless variables are constructed; $s=|\nabla n|/(2k_{\mathrm{F}}n)$ with $k_{\mathrm{F}}=(3\pi^2n)^{1/3}$ relevant for exchange, and $t_c=|\nabla n|/(2k_sn)$ with $k_s=\sqrt{4k_F/\pi}$ for correlation \cite{Perdew2006} measure the inhomogeneity of electron densities at the length scales of local Fermi wavelength $2\pi/k_{\mathrm{F}}$ and Thomas screening length $1/k_s$, respectively. Commonly used meta-GGAs include one more ingredient, the kinetic energy density $\tau(\vec r) = 1/2\sum^{\mr{occ.}}_i|\nabla \varphi_i(\vec r)|^2$ where $\{\varphi_i\}$ are KS orbitals. 

The inclusion of $\tau$ in meta-GGAs naturally arises from the Taylor expansion of the exact spherically averaged exchange hole \cite{Becke1983} and provides a simple and straightforward way to make a correlation functional exactly one-electron self-interaction free \cite{Becke1998}. The flexibility due to the inclusion of $\tau$ improves the accuracy of meta-GGAs over GGAs, by either better fitting to experimental data empirically \cite{VanVoorhis1998, Zhao2006, Mardirossian2015, Wang2017} or satisfying more exact constraints and appropriate norms non-empirically \cite{Tao2003, Perdew2009, Sun2012, Sun2015, Tao2016}. Early attempts \cite{Tao2003, Perdew2009} and the recent Tao-Mo functional \cite{Tao2016} construct non-empirical meta-GGAs through an iso-orbital indicator defined as,
\begin{equation}
    z = \frac{\tau^{\mr{vW}}}{\tau},
\end{equation}
where $\tau^{\mr{vW}}(\vec r) = |\nabla n(\vec r)|^2/8n(\vec r)$ is the von-Weisz\"acker kinetic energy density that recovers $\tau$ in the single orbital limit. Whilst $z$ can identify single orbital densities, $z = 0$, and slowly varying densities, $z \approx 1$, it is unable to distinguish slowly varying densities from the non-covalent closed shell overlap densities important in intermediate-range van-der-Waals bonding \cite{Sun2013a}. A different indicator widely used in empirical constructions \cite{Zhao2006,Peverati2012, Mardirossian2015} is,
\begin{equation}
    t^{-1} = \frac{\tau}{\tau^{\mr{UEG}}},
\end{equation}
where $\tau^{\mr{UEG}}=(3/10)(3 \pi ^2)^{2/3}n^{5/3}$ is the kinetic energy density of uniform electron gas. Whilst the $t^{-1}$ indicator has been shown to differentiate covalent from non-covalent bonding it cannot uniquely identify single-orbital regions for which $t^{-1} = 5s^{2/3}$. This is likely one of the major reasons for overfitting in the M06L meta-GGA and the resulting numerical stability problem \cite{Sun2013a}. $t^{-1}$ is semi-infinite, $[0,\infty]$, and usually mapped to a finite domain via $w=(1-t^{-1})/(1+t^{-1})=(\tau^{\mr{UEG}} -\tau)/ (\tau^\mr{UEG} +\tau)$.

A further iso-orbital indicator has been constructed,
\begin{equation}
    \alpha = \frac{\tau - \tau^{\mr{vW}}}{\tau^{\mr{UEG}}},
\end{equation}
which is able to uniquely identify single orbital, slowly varying and non-covalent overlap densities as $\alpha = 0$, $\approx 1$ and $\gg 1$ respectively. This indicator was included alongside $z$ in earlier meta-GGA functionals\cite{Tao2003, Perdew2009} to enforce the correct fourth-order gradient expansion\cite{Svendsen1996}, though its importance for characterising bonding densities was not fully recognised until later\cite{Sun2013a}. The $\alpha$ variable is directly related to the electron localization function (ELF) of Refs. \cite{Becke1990, Silvi1994} as $f_{\mr{ELF}} = 1/(1 + \alpha^2)$ which has been used to give a rigorous topological classification of chemical bonding \cite{Savin1996,Savin1997,Noury1998}. In addition, recent $\alpha$ dependent functionals\cite{Sun2015} have been found to effectively handle properties that have traditionally been challenging for semi-local functionals\cite{Paul2017,Zhang2017a, Zhang2018, Sun2016, Furness2018}, including the intermediate-range van-der-Waals bonding \cite{Sun2016} and metal-insulator transitions \cite{Furness2018}.

Despite the general success enjoyed by recent meta-GGA functionals for a wide range of systems\cite{Tao2003, Zhao2006, DelCampo2012, Sun2015, Mardirossian2015, Sun2016, Yang2016, Zhang2018, Furness2018}, it has been observed that many meta-GGAs suffer numerical instabilities in self-consistent field (SCF) calculations \cite{Johnson2009, Yang2016}, and have an unacceptably slow convergence with respect to the density of the numerical integration points \cite{Mardirossian2017a}. The increased computational cost of dense numerical grids severely limits the usefulness of many meta-GGA functionals and restricts the complexity of systems to which they can be applied. Additionally, the uncertainty in overall grid convergence necessitates a time-consuming validation that calculated properties are properly converged in grid density, placing an undesirable burden of expertise on the user.

Within $\alpha$-based functionals, it is understood that this numerical sensitivity originates from sharp oscillations in the exchange-correlation potential that are only properly captured by very fine grids\cite{Yang2016}, particularly in inter-shell regions where the local orbital overlap character can vary rapidly. Here we show that the rapid variations in the derivatives of $\alpha$ are largely responsible for these undesirable oscillations. Further, we propose a modified iso-orbital indicator quantity, termed $\beta$, from which new functionals can be constructed that do not suffer this problem,
\begin{align}
\beta(\vec r) &= \frac{\tau(\vec r) - \tau^{\mr{vW}}(\vec r)}{\tau(\vec r) + \tau^{\mr{UEG}}(\vec r)} \\
&= \alpha(\vec r) \left(\frac{\tau^{\mr{UEG}}(\vec r)}{\tau(\vec r) + \tau^{\mr{UEG}}(\vec r)}\right).
\end{align}
The $\beta$ variable contains similar information about local orbital overlap environments to $\alpha$ whilst having smoother derivatives more easily amenable to evaluation on numerical grids. This allows local orbital overlap information to be used in functionals at the meta-GGA level and higher without suffering the numerical problems associated with analogous functions of the $\alpha$ variable.

\begin{table}
\caption{\label{tab:alphabeta} Values of common iso-orbital indicators for important chemical environments.}
\begin{tabular}{lcccc}
\hline\hline
Region & $t^{-1}$ & $z$ & $\alpha$ & $\beta$ \\
\hline
Single orbital & $5s^2/3$ & 1 & $0$ & $0$ \\
Slowly varying density & $\approx 1$ & $\approx 0$ & $\approx 1$ & $\approx \frac{1}{2}$ \\
Non-covalent bonding & $\gg 1$ & $\approx 0$ & $\gg 1$ & $\frac{1}{2}  \ll \beta < 1$\\
\hline\hline
\end{tabular}
\end{table}

The close relationship between the $\alpha$ and $\beta$ variables can be shown by examining their limits in the three important orbital overlap regions. Firstly, in single orbital regions $\tau(\vec r) = \tau^{\mathrm{vW}}(\vec r)$ and $\alpha(\vec r) = \beta(\vec r) = 0$. This bound is important in exchange-correlation functional development, allowing the strong lower bound on exchange energy as well as correlation energy to be enforced for all spin-unpolarised single-orbital systems \cite{Perdew2014, Sun2015, Sun2015a}. Secondly, in slowly varying densities $\tau^{\mathrm{vW}}(\vec r) \rightarrow 0$ and $\tau(\vec r) \approx \tau^{\mr{UEG}}(\vec r)$, so $\alpha(\vec r) \approx 1$ and $\beta(\vec r) \approx 1/2$. Finally, in non-covalent density overlap regions where $n(\vec r) \rightarrow 0$, $\tau^{\mr{UEG}}(\vec r) \rightarrow 0$ as $n(\vec r)^{5/3}$ while $\tau(\vec r) - \tau^{\mathrm{vW}}(\vec r) \rightarrow 0$ as $n(\vec r)$, and thus the denominator, $\tau^{\mr{UEG}}(\vec r)$, decays faster than the numerator, $\tau(\vec r) - \tau^{\mathrm{vW}}(\vec r)$, leading to $\alpha(\vec r) \rightarrow \infty$. Here $\beta(\vec r)$ approaches $1$ however, since both its numerator and denominator decay as $n(\vec r)$. The relations between $\alpha$, $\beta$ and other iso-orbital indicators are summarised in Table \ref{tab:alphabeta}.

\begin{figure}
\includegraphics[width=\linewidth]{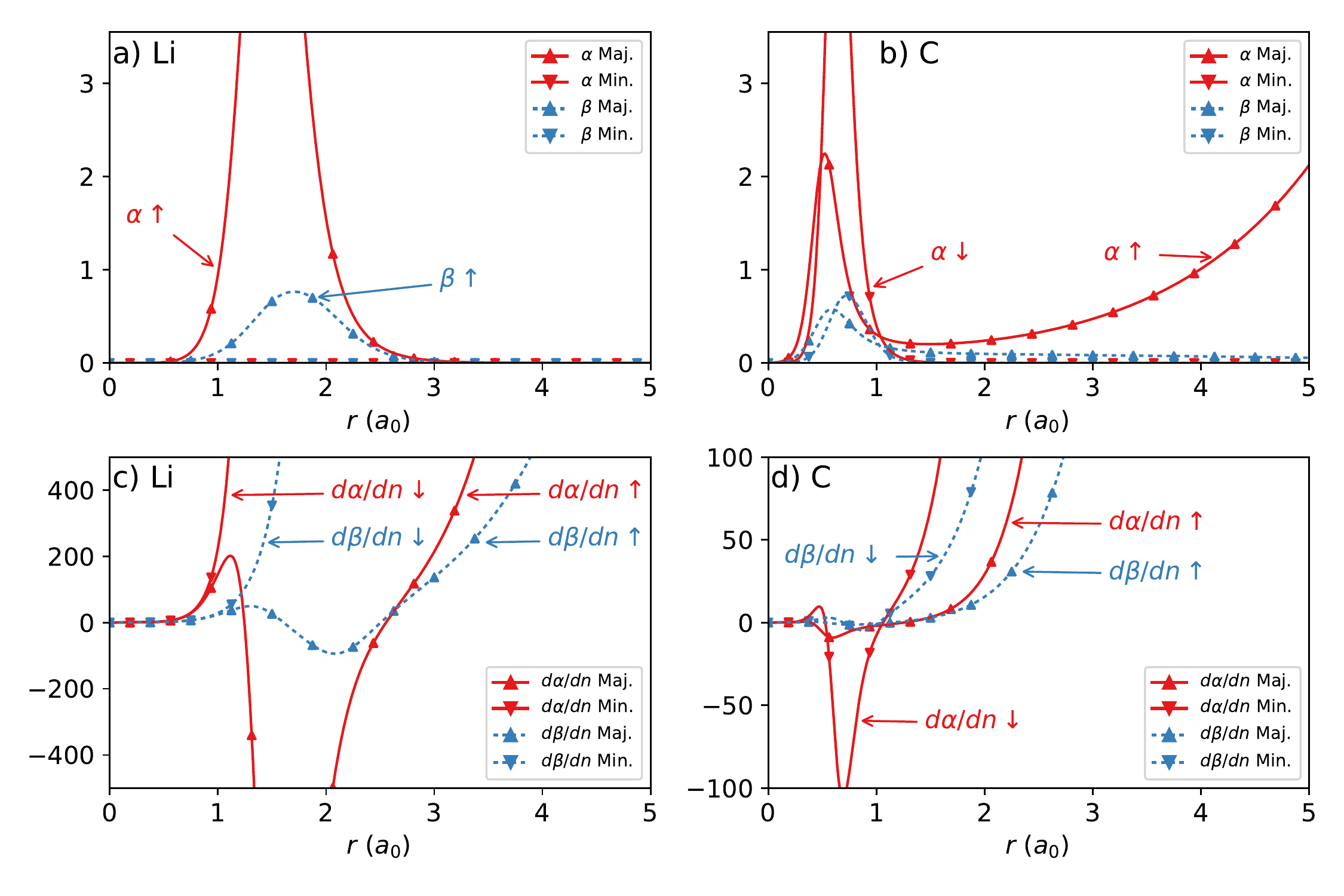}
\caption{\label{fig:derivs} \textbf{Radial plots of $\alpha(r)$ and $\beta(r)$ functions and their density derivatives for the lithium and carbon atoms.} The $\alpha$(solid red) and $\beta$(dashed blue) functions and density derivatives are plotted for the majority (Maj., $\blacktriangle$) and minority(Min., $\blacktriangledown$) spin from self-consistent PBE \cite{Perdew1996} densities. Radial distances are in units of Bohr.}
\end{figure}

This similarity between $\alpha$ and $\beta$ in highlighting local orbital overlap environment is shown graphically in Figures \ref{fig:derivs} a) and b) for the lithium and carbon atoms respectively. The lithium and carbon atoms were chosen as typically difficult systems with pronounced numerical sensitivity for density functionals. The similar inter-shell peak behaviour is clearly visible for both variables as a rise away from single orbital like values of the $1s$ core. The difference between $\alpha$ and $\beta$ is seen in the tail region of the carbon atom for the majority spin electrons, where $\alpha$ diverges upwards whilst $\beta$ decays slowly towards 0. For the minority spin, both $\alpha$ and $\beta$ approach 0 in the tail region. In general, $\beta$ consistently indicates the tail regions of all densities with 0 as $\tau^{\mathrm{vW}}(\vec r) $ is the leading order of $\tau(\vec r)$ there, and identifies tail regions completely when combined with the dimensionless reduced density gradient, $s$, that is divergent. This is impossible for $\alpha$ which can have different values of $\alpha(\vec r) = 0$ for the tail regions of single orbital systems and $\alpha \rightarrow \infty$ otherwise.

The benefit of $\beta$ over $\alpha$ for functional development is most clearly seen in the derivatives with respect to density, exemplified in Figure \ref{fig:derivs} c) and d) for the lithium and carbon atoms respectively. The derivative of $\beta$ with respect to density does not show the same rapid variation as that of $\alpha$ in valence regions where density is significant. Similar behaviour is observed for $|\nabla n|$ and $\tau$ derivatives, as noted for $\alpha$ in Ref. \cite{Yang2016} and in the more complex molecular systems presented in supplementary material Section 1. 

These much better behaved derivatives of $\beta$ over those of $\alpha$ are expected to remedy the numerical problems observed in $\alpha$-dependent meta-GGA functionals. To validate this, we modify the existing \emph{meta-GGA made simple 2} (MS2) functional \cite{Sun2013} by substituting $2\beta$ in place of $\alpha$ within the exchange enhancement factor $F_{\mathrm{x}}(s, \alpha)$, defined by
\begin{equation}
E_{\mathrm{x}}^{\mr{meta-GGA}}[n]=\int e_{\mathrm{x}}^{\mr{UEG}}(n)  F_{\mathrm{x}}^{\mr{meta-GGA}}(s, \alpha) d\mathbf{r},
\label{eq:Fx}
\end{equation}
where $e_{\mathrm{x}}^{\mr{UEG}}(n)=-3(3 \pi^2)^{1/3} n^{4/3}/4 \pi$ is the exchange energy density of uniform electron gas. MS2 has a simple construction of $F_{\mathrm{x}}^{\mr{MS2}}(s, \alpha) = F_x^1(s) + f_x^{\mr{MS2}}(\alpha) [ F_x^0(s) - F_x^1(s) ]$ with 
\begin{equation}
    f_{\mr{x}}^{\mr{MS2}}(\alpha) = \frac{\left(1 - \alpha^2\right)^3}{1 + \alpha^3 + b\alpha^6}
\end{equation}
that interpolates between $F_x^0(s)$, a GGA for single-orbital systems ($\alpha = 0$), and $F_x^1(s)$, a GGA for slowly-varying densities ($\alpha \approx 1$), and extrapolates to ($\alpha \gg 1$) for noncovalent bonds.

\begin{table}
\caption{\label{tab:params}Parameters for the MS2 and MS2$\beta$ functionals. The notation of Ref. \cite{Sun2013} is followed, with constants $k_0 = 0.174$ and $\mu_{\mathrm{GE}} = 10/81$. $^\dagger$The original publication of MS2\cite{Sun2013a} gives $c = 0.14601$. We find this gives a small error (on the order of $\mu E_{\mr{h}}$) in the exchange energy of the hydrogen atom that is corrected using $c = 0.14607$. We note however, that this change has a completely negligible impact on MS2 predicted atomisation energies and barrier heights.}
\begin{tabular}{llll}
\hline\hline
MS2  & $\kappa = 0.504$ & $b = 4.0$ & $c = 0.14607^\dagger$ \\
MS2$\beta$ & $\kappa = 0.504$ & $b = (27b^{\mathrm{MS2}} - 9)/64$ & $c = 0.14607$ \\
\hline\hline
\end{tabular}
\end{table} 

This functional was chosen for the relatively simple construction and well reported numerical instabilities \cite{Yang2016}. Modification of the simple exchange enhancement factor for $\beta$ dependence was trivial by replacing $\alpha$ with $2\beta$ to guarantee the interpolation between $F_x^0(s)$ for single orbital systems ($2\beta = \alpha = 0$) and $F_x^1(s)$ for slowly varying densities  ($2\beta \approx \alpha \approx 1$). The resulting $\beta$-modified meta-GGA functional is termed MS2$\beta$. The parameter $b$ is determined such that $f_{\mr{x}}^{\mr{MS2\beta}}(\beta=1)=f_{\mr{x}}^{\mr{MS2}}(\alpha \to \infty)$ for noncovalent bonds. These minor adjustments to the balances between internal parameters to preserve exact constraints obeyed by the parent functional are summarised in Table \ref{tab:params} and detailed in supplementary material. MS2$\beta$ was implemented into the \textsc{Turbomole} package \cite{TURBOMOLE} using the \textsc{XCFun} library \cite{Ekstrom2010} to automatically evaluate functional derivatives. We present MS2$\beta$ simply as a convenient means for preliminary investigation of the $\beta$ variable rather than as a viable general functional.

\begin{figure}
\includegraphics[width=\linewidth]{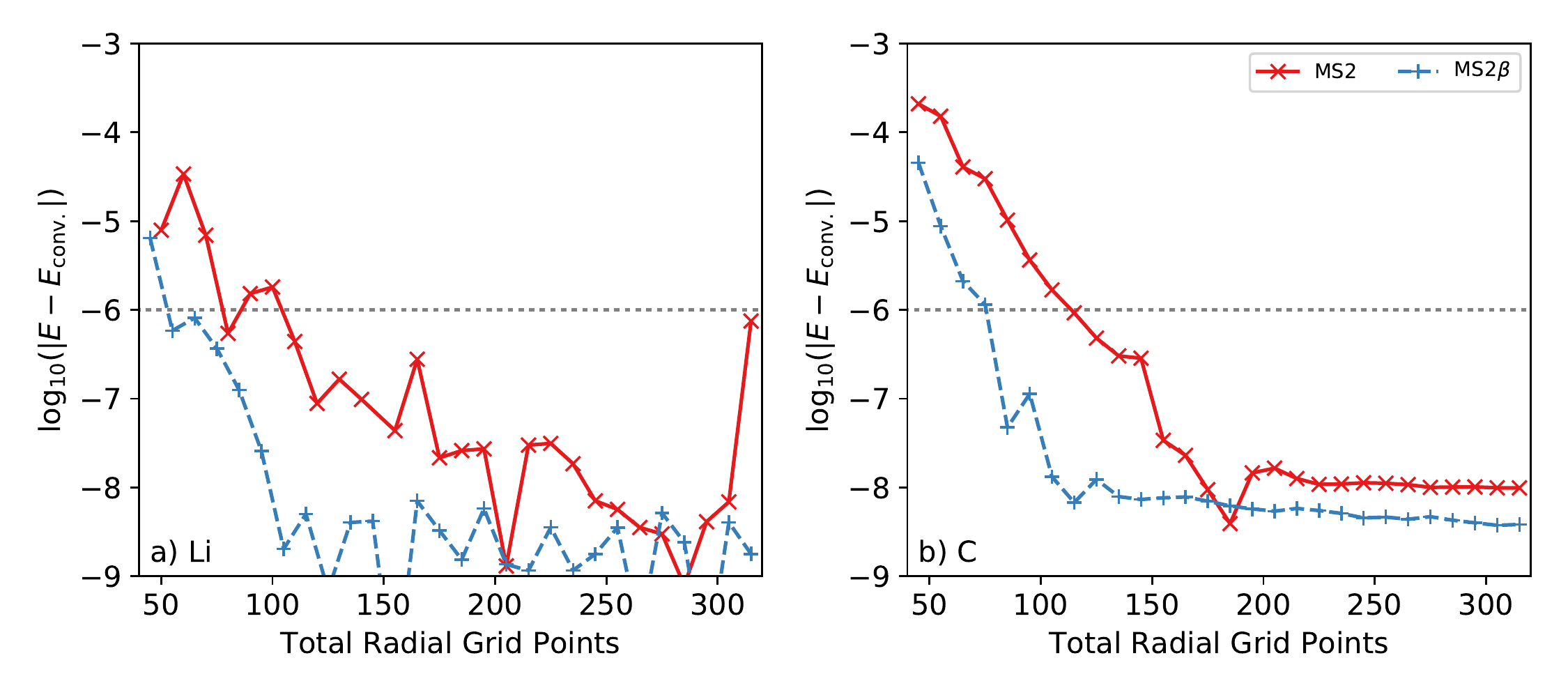}
\caption{\label{fig:gridconv} \textbf{Convergence behaviour with respect to numerical grid density of conventional and $\beta$-modified \emph{meta-GGA made simple 2} (MS2) functional\cite{Sun2013} for (a) lithium and (b) carbon atoms.} Difference in self-consistent atomic total energy relative to the total energy from a converged grid of 520 points ($E_{\mathrm{conv.}}$) for the MS2 (red, $\times$) and MS2$\beta$ (blue, dashed, $+$) functionals using the aug-cc-pVQZ basis set \cite{Dunning1989}. An acceptable convergence is assumed when difference remains below the SCF convergence tolerance of 1$\mu E_{\mathrm{h}}$ (dotted line).}
\end{figure}

The grid convergence of the MS2 and MS2$\beta$ functionals is shown in Figure \ref{fig:gridconv} as a plot of the difference in self-consistent electronic energy relative to the same calculation using a very fine benchmark grid, as a function of grid point density, for the carbon and lithium atoms. In both cases the $\beta$-modified functional shows a convergence in total energy at lower grid density than the parent $\alpha$ functional, indicated by the difference to the benchmark grid energy remaining below the SCF convergence threshold of 1$\mu E_h$.

As previously noted, the sensitivity of $\alpha$-dependent functionals to numerical integration grid point density is understood as an effect of rapid oscillations in the exchange-correlation potential, expressed in the generalised Kohn--Sham scheme\cite{Neumann1996, Seidl1996, Yang2016} as,
\begin{equation}
\int \varphi_p \hat v_{\mathrm{xc}} \varphi_q d\mathbf{r} = \int \varphi_p \frac{\partial e_{\mathrm{xc}}}{\partial n} \varphi_q d\mathbf{r} + \int \frac{\partial e_{\mathrm{xc}}}{\partial \nabla n} \cdot \nabla(\varphi_p \varphi_q) d \mathbf{r} + \frac{1}{2}\int \nabla \varphi_p\cdot \left(\frac{\partial e_{\mathrm{xc}}}{\partial \tau} \right) \nabla \varphi_q d\mathbf{r}
\label{eq:gKS}
\end{equation}
Here, $e_{\mathrm{xc}}$ is the exchange-correlation energy density of a meta-GGA functional.
Hence, functionals showing sharp oscillations in functional derivatives will show sharp variations in exchange-correlation potential that are challenging to capture in numerical integration schemes. By modifying meta-GGA functionals to use the $\beta$ indicator in place of $\alpha$, oscillations in functional derivatives are minimised and the resulting exchange-correlation potential is smoothed, as exemplified by MS2 and MS2$\beta$ which have the same exchange enhancement factor form in terms of $\alpha$ or $2\beta$ dependence. We can therefore understand the reduced grid sensitivity of the $\beta$-modified functional as a consequence of smoother functional derivatives in the inter-shell and valence regions producing smoother exchange-correlation potentials that can be more easily captured by numerical integration grids.

The divergences of derivatives at tails seen in Figures \ref{fig:derivs} c) and d) are not problematic for the exchange-correlation potential of any analytic $\beta$-dependent meta-GGAs because $e_{\mr{x}}^{\mr{UEG}}(n)$ decays faster than the divergence of $\beta$ derivatives there. However, the divergences of $\alpha$ derivatives at tails are much stronger than those of $\beta$, as also seen in Figures \ref{fig:derivs} c) and d), and can cause problems for the exchange-correlation potential of $\alpha$-dependent meta-GGAs. At the tail regions, $\tau$ decays as $n$, whilst $\tau^{\mr{UEG}}$ decays more quickly as $n^{5/3}$, so the derivative of $\alpha$ with respect to, for example, $\tau$ diverges as $n^{-5/3}$, faster than the decay of $e_{\mr{x}}^{\mr{UEG}}(n)$ proportional to $n^{4/3}$. This can potentially lead to divergences of exchange-correlation potential for $\alpha$-dependent meta-GGAs as long as the derivative of $F_{\mr{x}}^{\mr{meta-GGA}}(s, \alpha)$ with respect to $\alpha$ is non-zero at the tail. Such non-zero values are encountered by the \emph{meta-GGA made very simple} (MVS) \cite{Sun2015a} and SCAN \cite{Sun2015} meta-GGAs at the tail regions of single-orbital systems. This behaviour is avoided in $\beta$ as the decay of both the numerator and denominator is determined by $\tau$ as decaying with $n$.

The potential divergence of exchange-correlation potential at tail regions resulting from the strong divergence of $\alpha$ derivatives is undesirable and problematic especially for the construction of pseudo-potentials from isolated atoms, acting as a hitherto unrecognised additional constraint on $\alpha$ dependent functional design. This constraint was not previously noticed and is not obeyed by the MVS \cite{Sun2015a} and SCAN \cite{Sun2015} functionals, though it is fortuitously obeyed by MS2. This constraint is not necessary in $\beta$ dependent functionals however, as the asymptotic behaviour of $\beta$ is properly controlled and thus $\beta$ offers simplicity and greater flexibility in functional design compared to the $\alpha$ indicator.

\begin{table}
\caption{\label{tab:ae6bh6} Mean error (ME) and mean absolute error (MAE) in kcal mol$^{-1}$ for the small test sets of atomisation energy (AE6) and reaction barrier height (BH6) for the MS2 and MS2$\beta$ functionals. All calculations were performed fully self-consistently with the 6-311++G(3df,3pd) basis set\cite{Clark1983, Frisch1984a} on a dense numerical grid (\textsc{Turbomole} level 7).}
\begin{tabular}{l|rr}
\hline\hline
 & MS2 & MS2$\beta$ \\
\hline
AE6 (ME)  & -0.78 &  3.95 \\
AE6 (MAE) &  4.40 &  6.10 \\
BH6 (ME)  & -5.79 & -6.32 \\
BH6 (MAE) &  5.79 &  6.32 \\
\hline\hline
\end{tabular}
\end{table}

\begin{figure}
\includegraphics[width=\linewidth]{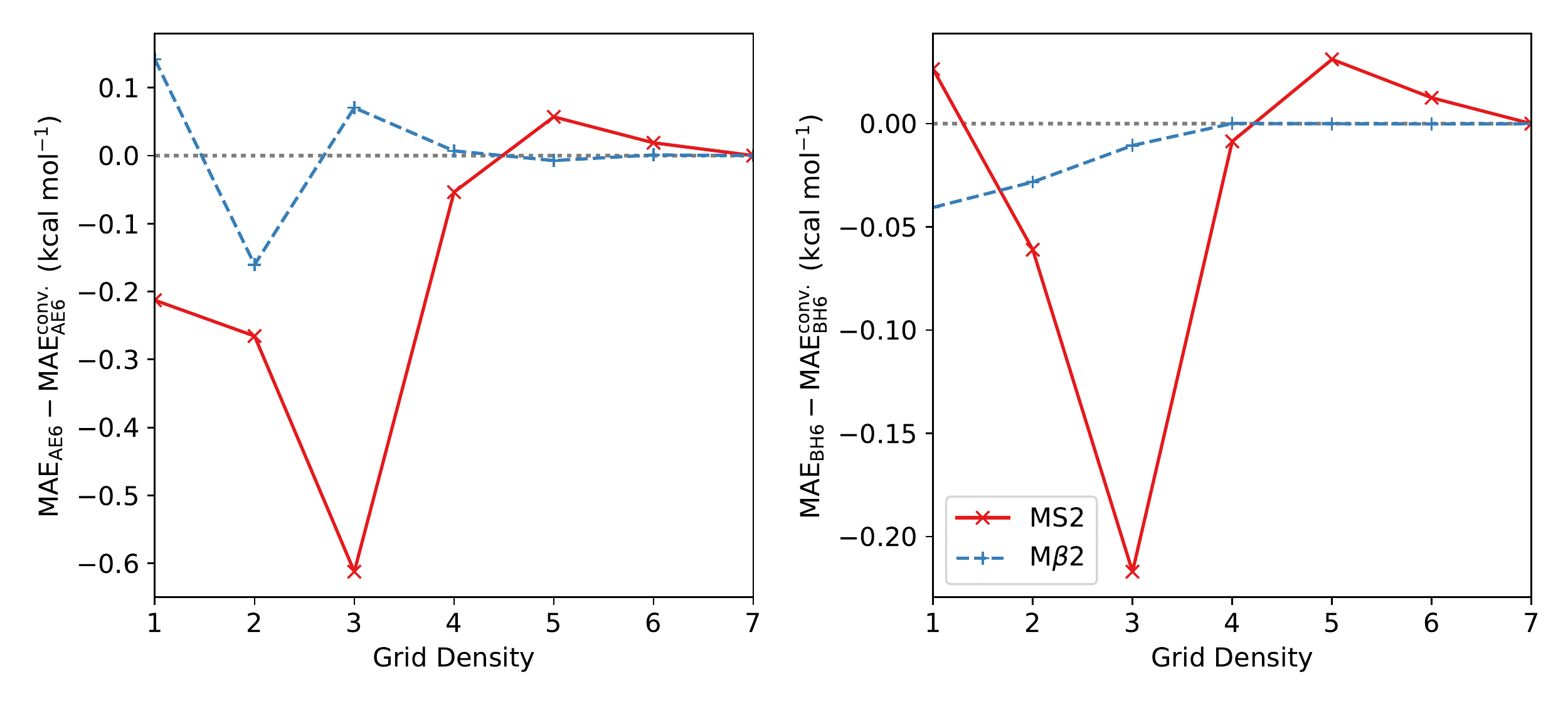}
\caption{\label{fig:setconv} Convergence of mean absolute error for the AE6 and BH6 small molecule data sets \cite{Lynch2003, Haunschild2012}, with respect to increasing numerical grid density for the MS2 (red, $\times$) and MS2$\beta$ (blue, dashed, $+$) functionals. The presence of both angular and radial grid components and the variety of molecules in the sets prevents a simple quantitative measure of grid density. As such, grid density is reported as the chosen \textsc{Turbomole} grid level, such that a higher grid level (larger $x$ axis value) directly corresponds to more dense integration grids, though steps between levels may not be uniform. All calculations were performed fully self-consistently with the 6-311++G(3df,3pd) basis set\cite{Clark1983, Frisch1984a}.}
\end{figure}

To show that the MS2$\beta$ functional retains the ability to effectively distinguish different chemical environments we calculate the mean error (ME) and mean absolute error (MAE) of small data sets of atomisation energies (AE6) and barrier heights (BH6) \cite{Lynch2003, Haunschild2012}, the results for which are shown in Table \ref{tab:ae6bh6}. For these small datasets the accuracy of the $\beta$-modified functional is slightly reduced compared to the original. This is to be expected as the MS2 functional contains two free parameters fit to minimise MAE in these data sets that may not be optimal for $\beta$. We conclude therefore that whilst it is possible to use $2\beta$ directly in place of $\alpha$ in existing functionals, some adjustment of internal parameters and interpolation-extrapolation functions is likely to be necessary to construct useful general $\beta$ dependent functionals.

A comparison of the data set grid convergence for the MS2 and MS2$\beta$ functionals is summarised in Figure \ref{fig:setconv} as a comparison of increasing grid density with respect to the densest grid. A similar trend to the convergence of atomic total energies, Figure \ref{fig:gridconv}, is seen with MS2$\beta$ reaching a converged error at coarser numerical grids than the parent MS2 functional. We note that whilst convergence for the BH6 set is relatively rapid for both functionals much greater variation is seen for the atomisation energies in the AE6 set, consistent with the larger energy scale of the test set.

\begin{figure}
\includegraphics[width=0.5\linewidth]{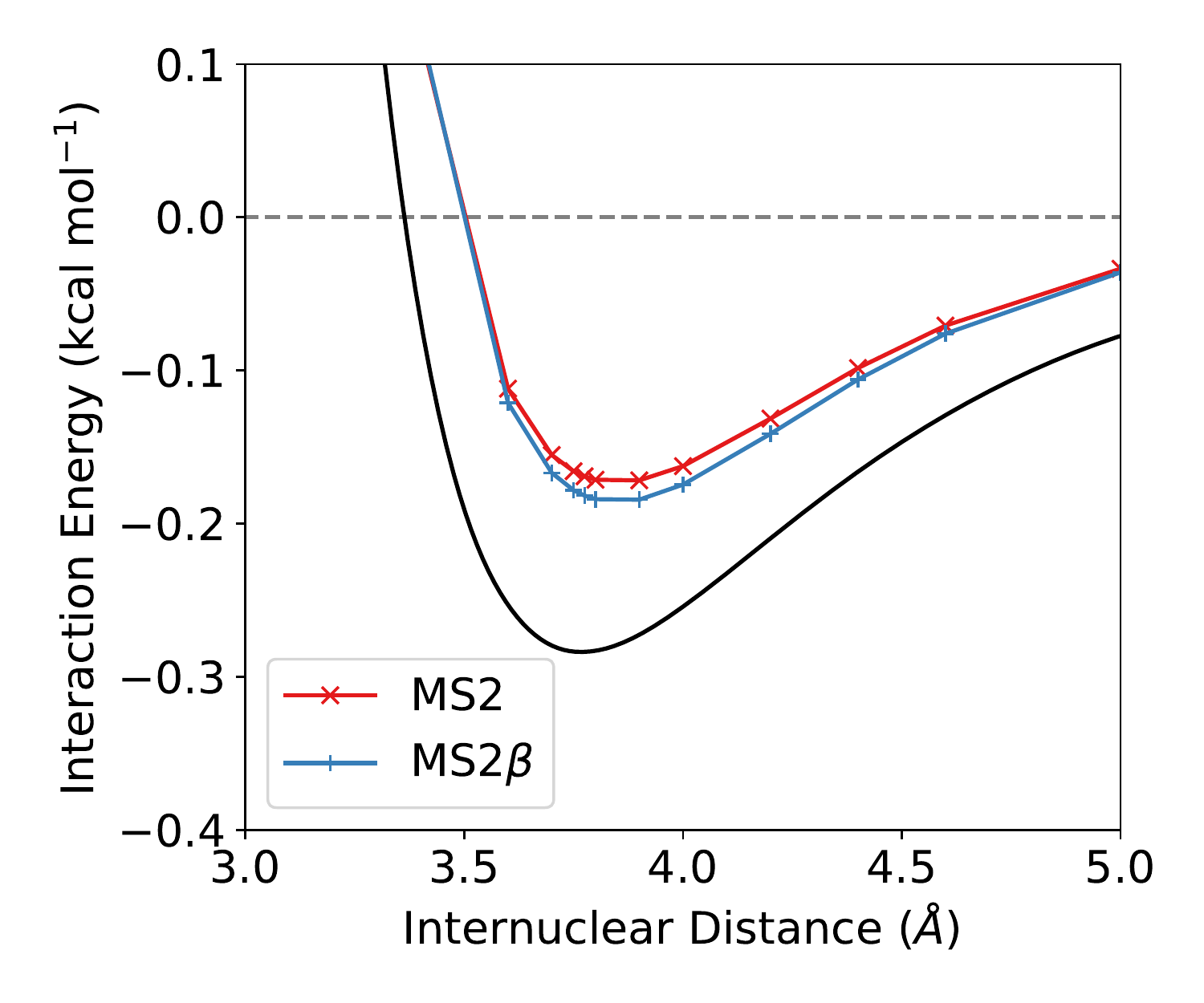}
\caption{\label{fig:ar2} \textbf{Dissociation curve of the argon dimer.} Calculated using MS2 (red, $\times$) and MS2$\beta$ (blue, $+$) functionals. Calculations performed self-consistently using the aug-cc-pV5Z basis set and a very fine numerical grid (grid 7 of the \textsc{Turbomole} program). A benchmark curve (black) is included from Ref. \cite{Patkowski2005}.}
\end{figure}

The different behaviour of the $\alpha$ and $\beta$ variables is most pronounced in regions of non-bonded density overlap where $\alpha(\vec r) \gg 1.0$ and $\beta$ is in the region $0.5 < \beta(\vec r) < 1.0$. This difference is most clearly examined in the Ar\mol{2} dissociation curve, for which it has been shown meta-GGA functionals can be accurate around equilibrium \cite{Sun2013,Sun2015,Sun2016}. The dissociation curves for Ar\mol{2} are shown in Figure \ref{fig:ar2} for both the conventional and $\beta$-modified MS2 functionals with benchmark data included from Ref. \cite{Patkowski2005} for comparison. In contrast to the small test sets, MS2$\beta$ closely matches the performance of the original MS2 functional across the whole of the binding curve, suggesting that the MS2 interpolation-extrapolation function may be as appropriate for $2\beta$ as it is for $\alpha$ in non-covalently bound systems.


In conclusion, we have identified the source of the numerical sensitivities commonly suffered in calculations employing meta-GGA functionals as resulting from sharp oscillations in the functional derivatives of the commonly employed dimensionless variable $\alpha$. We have addressed these sensitivities by constructing a related dimensionless variable, $\beta$, which imparts similar information about local orbital overlap environment whilst having smoother functional derivatives. The enhanced numerical stability, improved recognition of tail regions, and freedom from constraints on exchange-correlation potential presented by $\beta$ are an appealing opportunity for enhancing the numerical performance of future functionals for both the meta-GGA level and higher level fully non-local functionals.

We have used the simple MS2 meta-GGA functional to show a simple proof of concept by substituting $2\beta$ for $\alpha$ in the functional with minimal adjustment of internal parameters. We find improved numerical performance in all cases, with the new functional giving converged properties from much coarser integration grids than the original MS2 functional. Whilst the MS2$\beta$ functional preserves much of the good accuracy of its progenitor, a slight degradation of accuracy is observed for small molecule test sets though performance is preserved for the non-covalently bound Ar\mol{2} diatomic. This suggests a need for the underlying functional to be adjusted for $\beta$ rather than naive substitution if good performance for appropriate norms is to be preserved alongside an adherence to exact constraints. We are optimistic that wholly novel functionals utilising the $\beta$ iso-orbital indicator can provide ever greater accuracy and utility for the wider DFT community.

\section{Acknowledgements}

We thank Jefferson Bates for his invaluable technical advice in running and modifying the \textsc{Turbomole} package. We also thank John Perdew, Adrienn Ruzsinszky, Mark Pederson, Weitao Yang, Arun Bansil, Samuel Trickey, Susi Lehtola and Albert Bartok-Partay for their fruitful discussions around the ideas presented here. The work at Tulane University was supported by the start-up funding from Tulane University, and by the U.S. DOE, Office of Science, Basic Energy Sciences grant number DE-SC0235021(core research). Calculations were carried out on the Cypress cluster at Tulane University, and we thank Dr. Hoang Tran, Dr. Carl Baribault, and Dr. Hideki Fujioka for their computational support

\providecommand{\latin}[1]{#1}
\makeatletter
\providecommand{\doi}
  {\begingroup\let\do\@makeother\dospecials
  \catcode`\{=1 \catcode`\}=2 \doi@aux}
\providecommand{\doi@aux}[1]{\endgroup\texttt{#1}}
\makeatother
\providecommand*\mcitethebibliography{\thebibliography}
\csname @ifundefined\endcsname{endmcitethebibliography}
  {\let\endmcitethebibliography\endthebibliography}{}

\clearpage



\setcounter{equation}{0}
\setcounter{figure}{0}
\setcounter{table}{0}
\part*{Supplementary Material}

\section{Behaviour of $\alpha$ and  $\beta$ and their derivatives in a range of chemical environments}

The $\alpha$ and $\beta$ quantities and their $d/dn$, $d/d|\nabla n|^2$ and $d/d\tau$ derivatives are plotted for atomic and molecular systems, chosen to illustrate a range of chemically diverse systems. All evaluated from self consistent Perdew-Burke-Ernzerhof (PBE) densities \cite{Perdew1996}.

A common scheme of $\alpha$ (blue) and $\beta$ (red, dotted) for majority ($\blacktriangle$) and minority ($\blacktriangledown$) spin channels is used for all plots.

\begin{figure}[h]
\includegraphics[width=\linewidth]{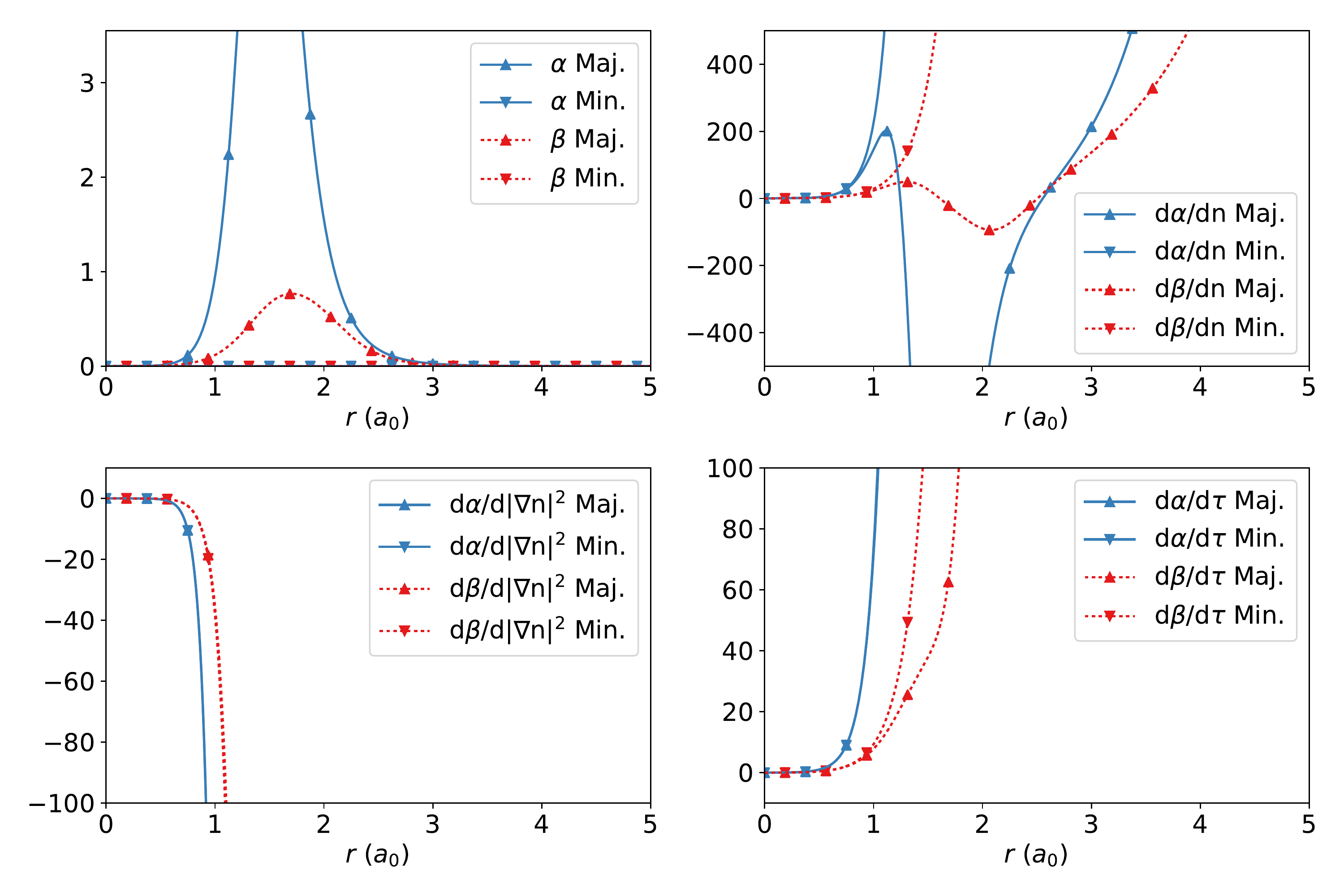}
\caption{\textbf{Radial plots of the $\alpha$ and $\beta$ iso-orbital indicators for the lithium atom.}}
\end{figure}

\begin{figure}[h]
\includegraphics[width=\linewidth]{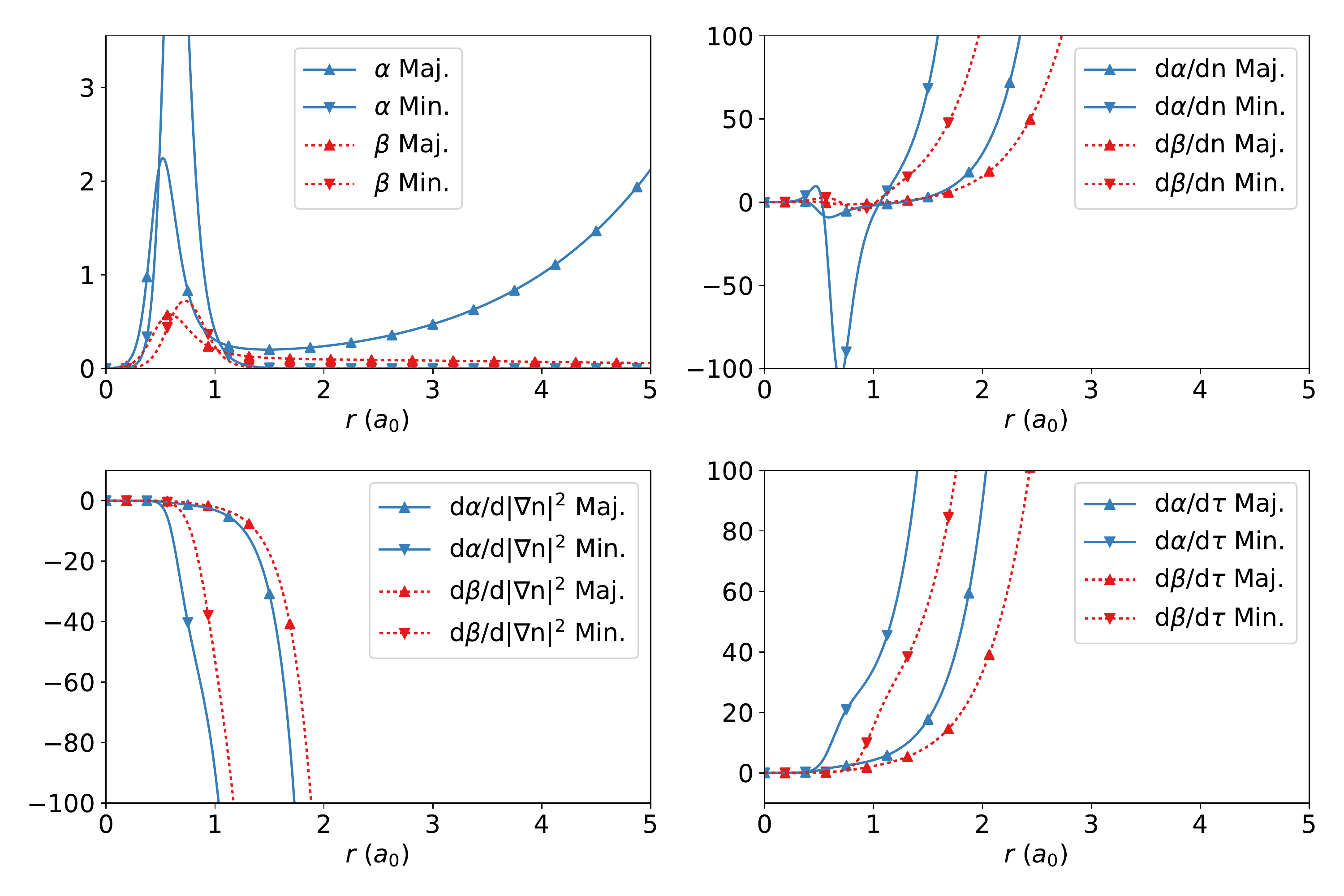}
\caption{\textbf{Radial plots of the $\alpha$ and $\beta$ iso-orbital indicators for the carbon atom.}}
\end{figure}

\begin{figure}[h]
\includegraphics[width=\linewidth]{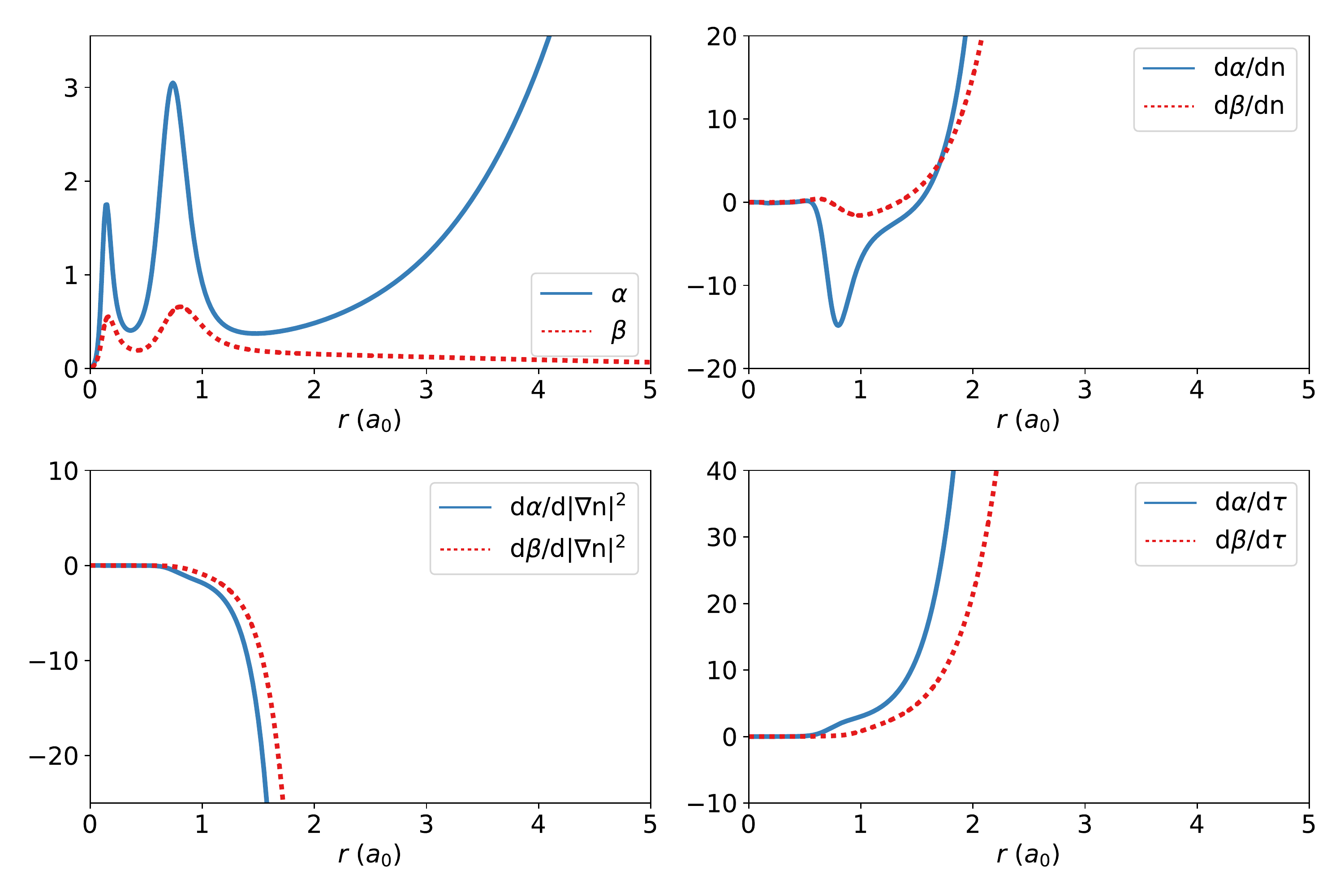}
\caption{\textbf{Radial plots of the $\alpha$ and $\beta$ iso-orbital indicators for the argon atom.}}
\end{figure}

\begin{figure}[h]
\includegraphics[width=\linewidth]{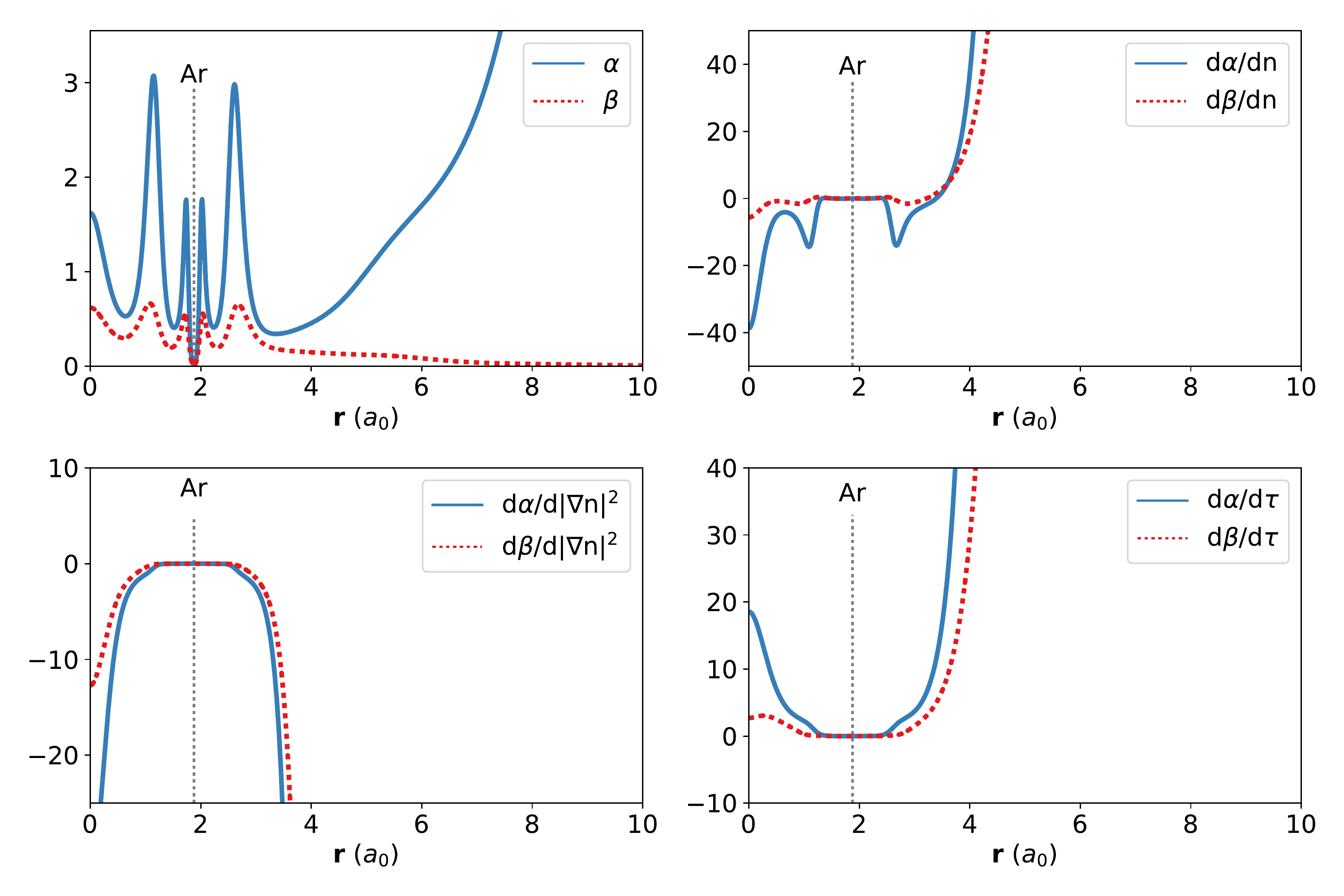}
\caption{\textbf{The $\alpha$ and $\beta$ iso-orbital indicators for the Ar\mol{2} vand-der-Waals diatomic, plotted along the internuclear axis evaluated for self consistent PBE density.} The origin is positioned at the bond center and argon atom position is marked with a dotted vertical line.}
\end{figure}

\begin{figure}[h]
\includegraphics[width=\linewidth]{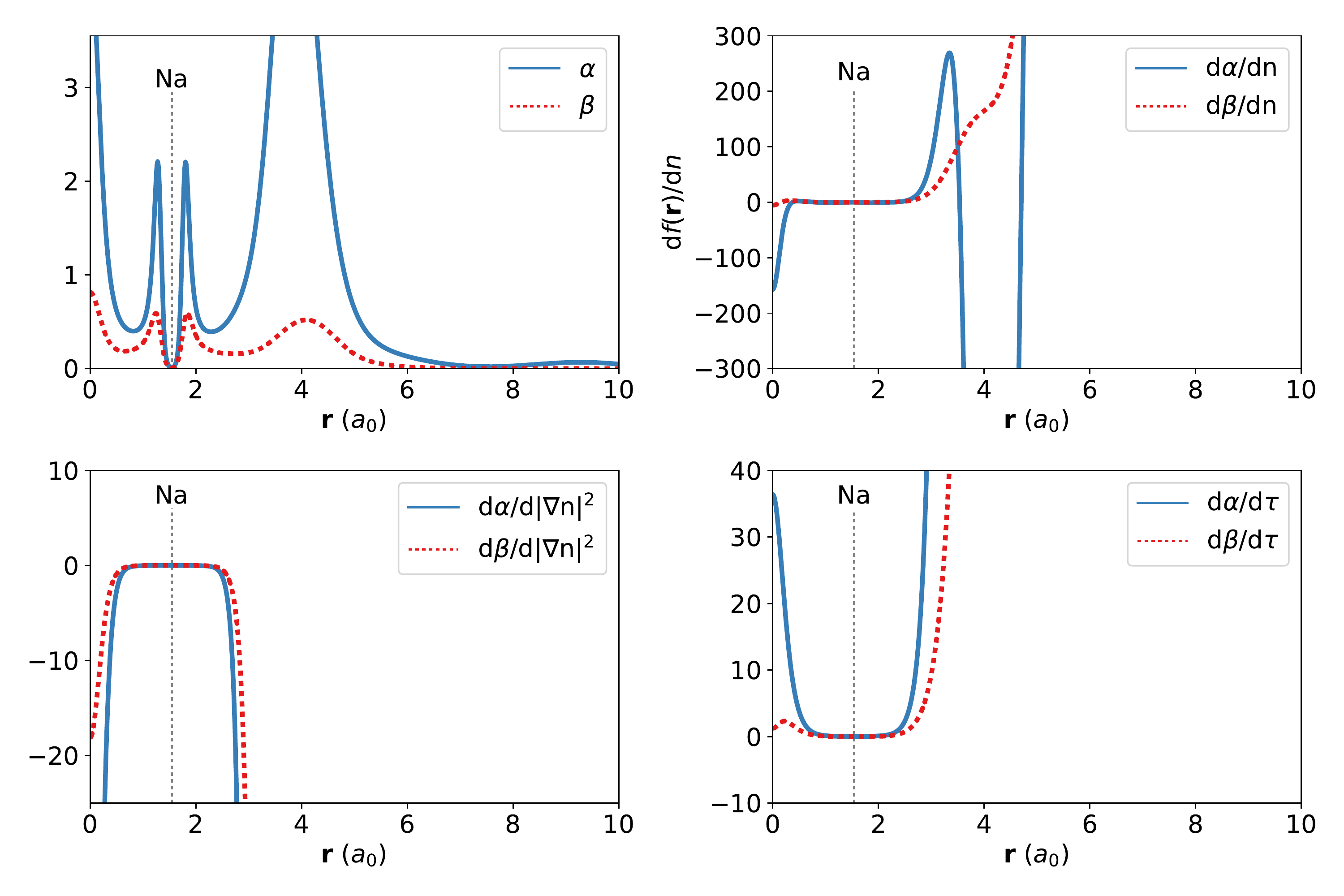}
\caption{\textbf{The $\alpha$ and $\beta$ iso-orbital indicators for the Na\mol{2} diatomic showing metallic bonding, plotted along the internuclear axis.} The origin is positioned at the bond center and sodium atom position is marked with a dotted vertical line.}
\end{figure}

\begin{figure}[h]
\includegraphics[width=\linewidth]{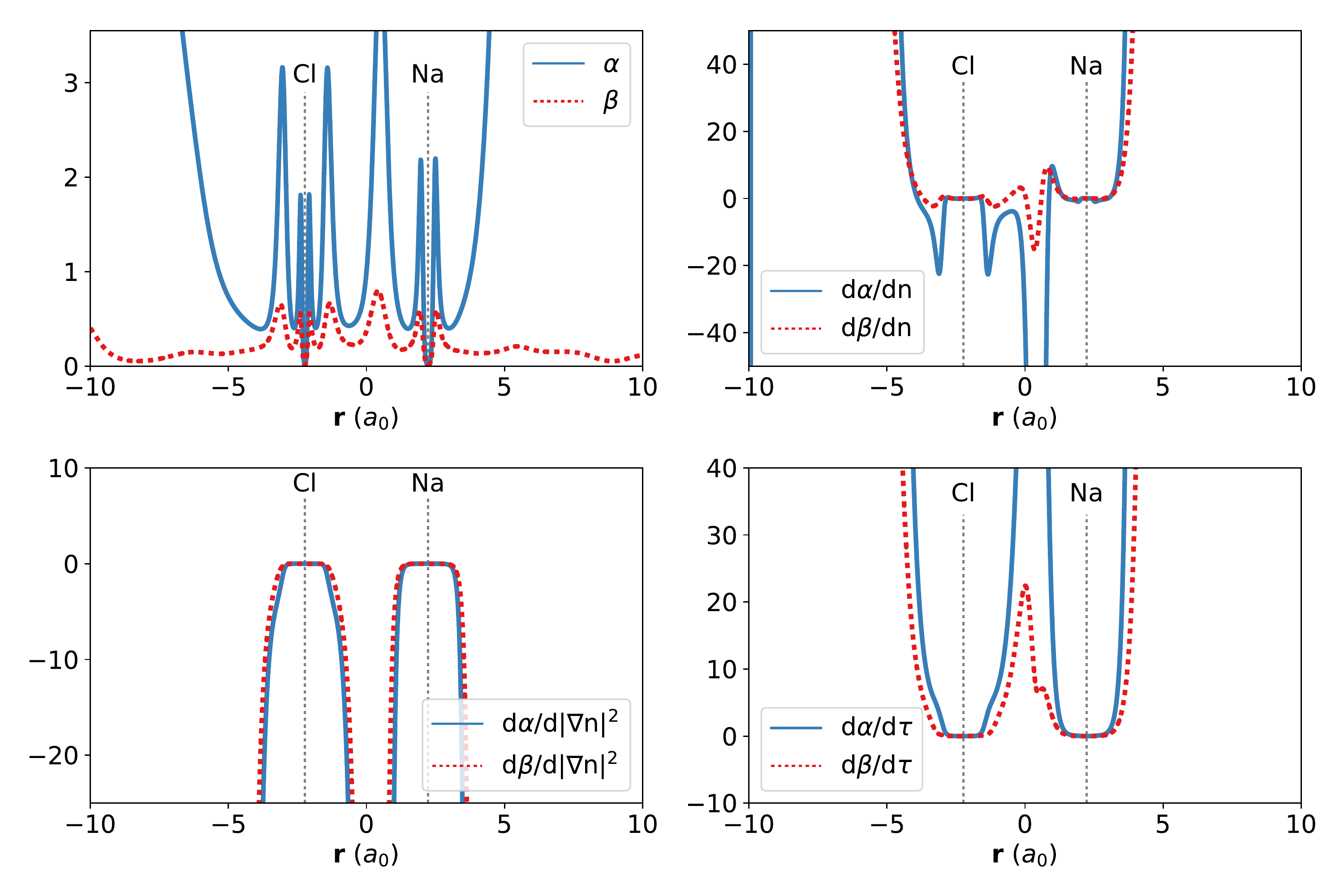}
\caption{\textbf{The $\alpha$ and $\beta$ iso-orbital indicators for NaCl showing ionic bonding, plotted along the internuclear axis.}}
\end{figure}

\begin{figure}[h]
\includegraphics[width=\linewidth]{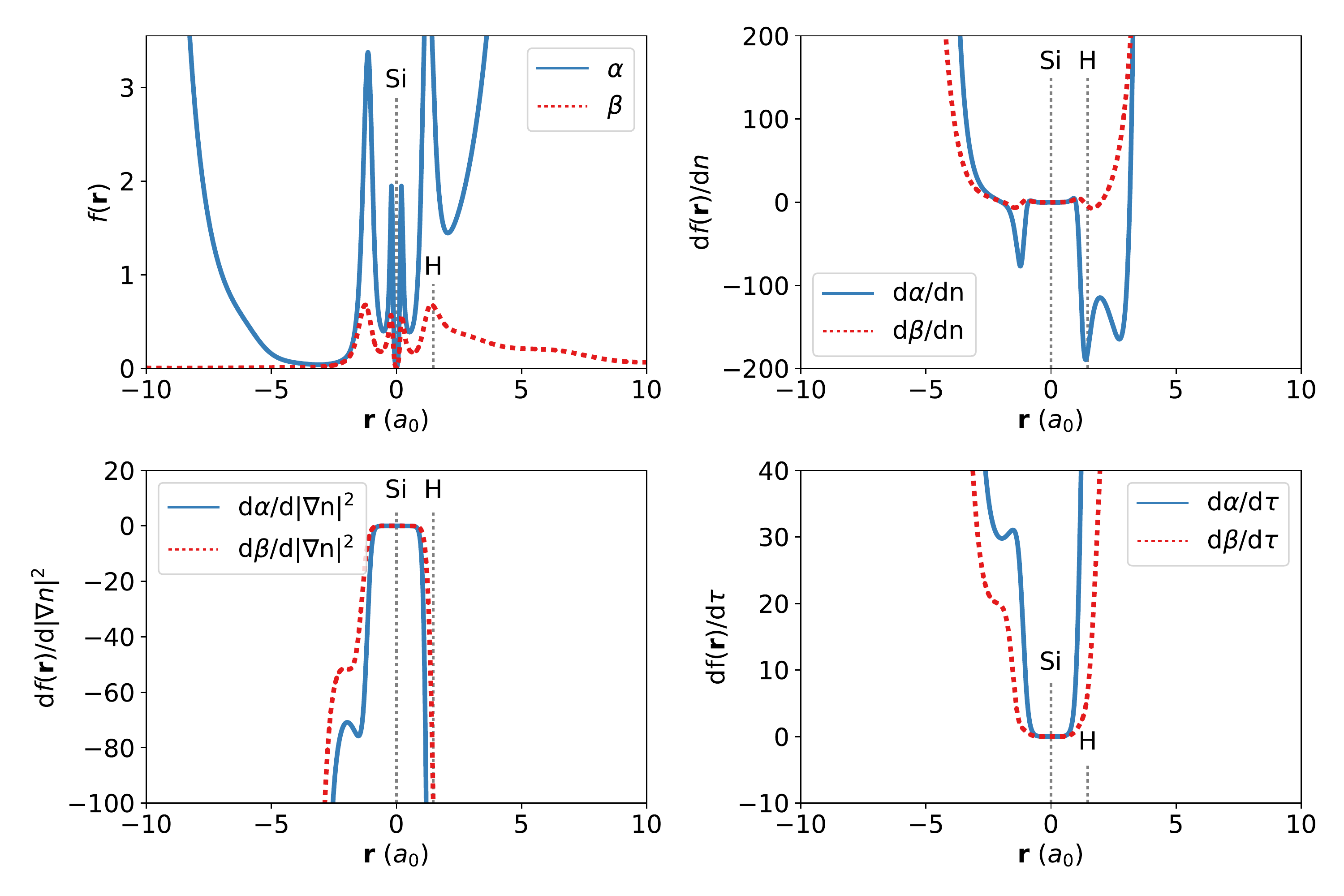}
\caption{\textbf{The $\alpha$ and $\beta$ iso-orbital indicators for SiH\mol{4} showing covalent bonding, plotted along one Si-H internuclear axis.}}
\end{figure}

\begin{landscape}
\begin{figure}[h]
\includegraphics[width=0.49\linewidth]{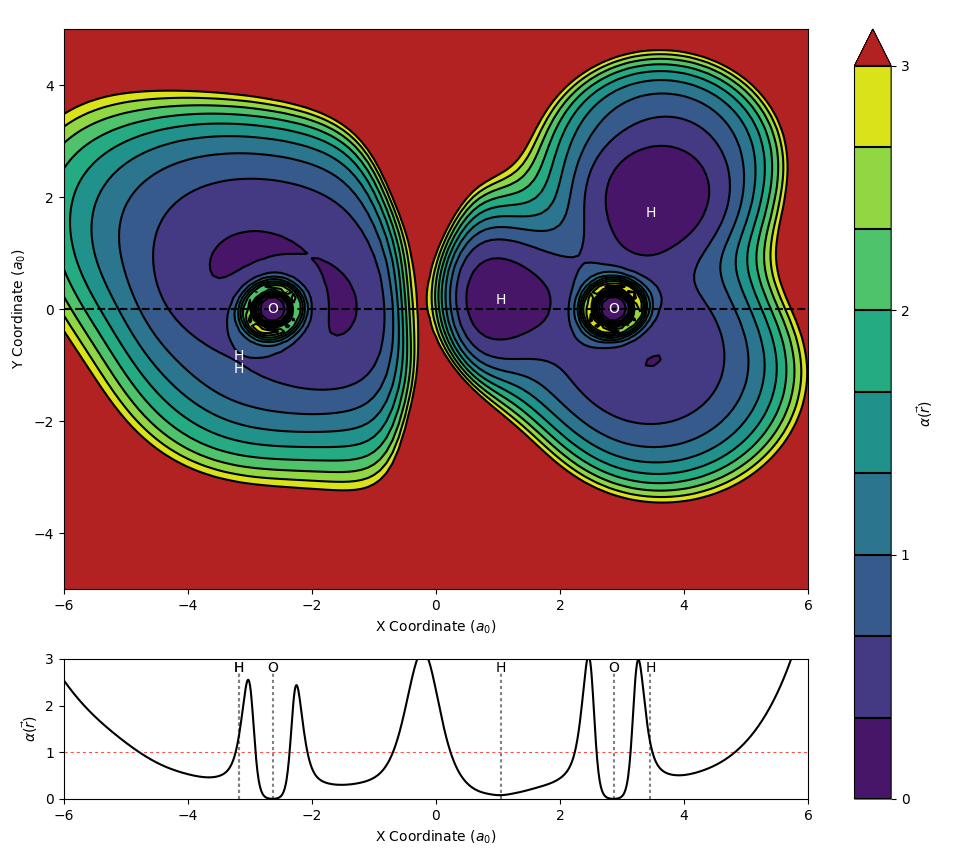}
\includegraphics[width=0.49\linewidth]{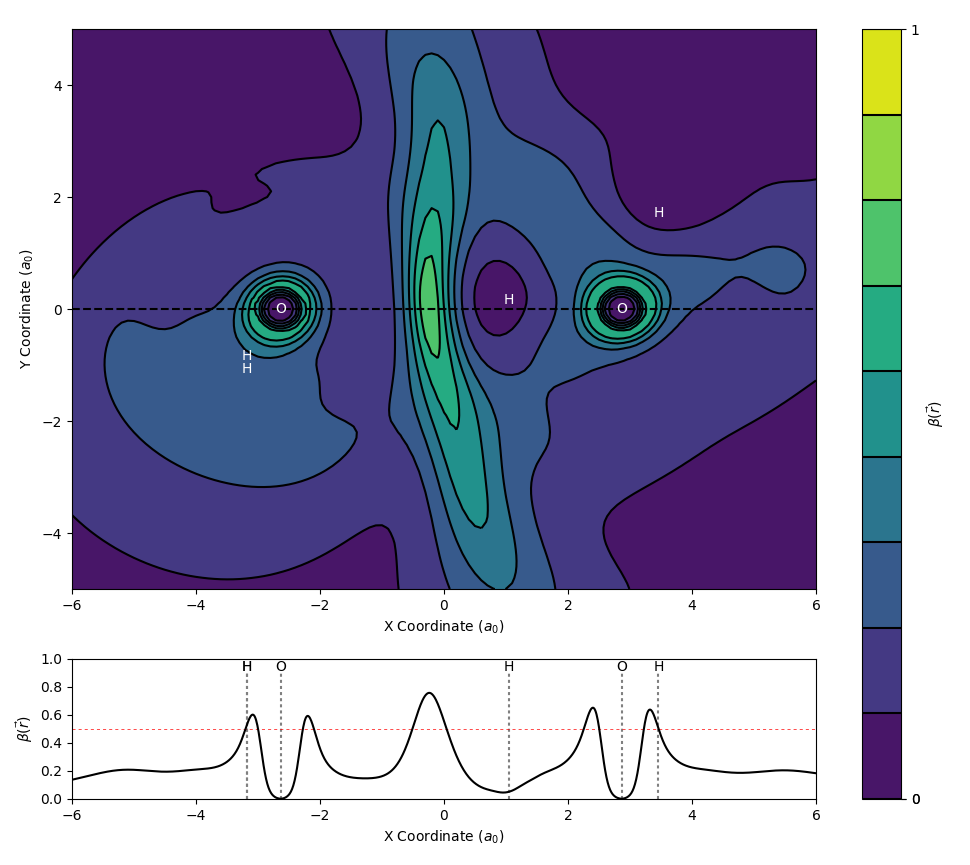}
\caption{\textbf{Contour plots of $\alpha(\vec r)$ (left) and $\beta(\vec r)$ (right) for the hydrogen bonded (H\mol{2}O)\mol{2} water dimer evaluated for self consistent PBE density.} Atomic positions are labelled, the labels for out of plane hydrogen atoms have been projected onto the plotting plane. As the range of the functions is different, colour mapping is not shared between plots. Additionally, as $\alpha(\vec r)$ is unbounded from above regions for which $\alpha(\vec r) > 3.0$ have been coloured red.
Linear slices along the inter-molecular axis (dotted black line on contour plot) are included, in which the slowly varying limits of 1.0 and 0.5 for $\alpha$ and $\beta$ respectively are marked with a red doted line.
Geometry is taken from the S22 test set \cite{Jurecka2006}, optimised at the B3LYP level \cite{Becke1993}.}
\end{figure}
\end{landscape}

\clearpage
\section{Derivation of the MS2$\beta$ functional}
The semilocal exchange functional can be written as a construction around an exchange enhancement factor modulating the exchange energy density of the uniform electron gas,
\begin{equation}
E_{\mr{x}}[n] = \int e_{\mr{x}}^{\mr{UEG}}(n)F_{\mr{x}}(s, \alpha) d\vec r,
\label{eq:mvsxc}
\end{equation}
where $e_{\mr{x}}^{\mr{UEG}}(n) = -\frac{3}{4}\left(\frac{3}{\pi}\right)^{1/3}n(\vec r)^{4/3}$ and $F_{\mr{x}}$ is the exchange enhancement factor.

The MS2 functional is designed to recover the second order gradient expansion of exchange and correlation energies of slowly varying densities.
For slowly varying densities, 
\begin{align}
\lim_{|\nabla n|\rightarrow 0}\alpha &= \frac{\tau^{\mr{UEG}} + \frac{1}{72}\frac{|\nabla n|^2}{n} + \frac{1}{6}\nabla^2 n - \frac{|\nabla n|^2}{8 n}}{\tau^{\mr{UEG}}} \\
&= 1 - \frac{1}{9} A + \frac{1}{6} B + O(\nabla^2), \label{eq:limalpha}
\end{align}
where,
\begin{align}
A &= \frac{|\nabla n|^2}{e_{\mr{x}}^{\mr{UEG}}}, \\
B &= \nabla^2 n/\tau^{\mr{UEG}},\label{eq:B}
\end{align}

Here, a similar derivation is made for the $\beta$ quantity,
\begin{align}
\lim_{|\nabla n| \rightarrow 0} \beta &= \frac{\tau^{\mr{UEG}} + \frac{1}{72}\frac{|\nabla n|^2}{n} + \frac{1}{6}\nabla^2 n - \frac{|\nabla n|^2}{8 n}}{\tau^{\mr{UEG}} + \frac{1}{72}\frac{|\nabla n|^2}{n} + \frac{1}{6}\nabla^2 n + \tau^{\mr{UEG}}}, \\
 &= \frac{1}{2}\left[1 - \frac{17}{144}A + \frac{1}{12}B\right] + O(\nabla^2).\label{eq:betaexp}
\end{align}

The MS2 exchange enhancement function, $F^{\mr{MS2}}_{\mr{x}}(p, \alpha)$ is constructed as an interpolation, $f(\alpha)$, of two enhancement factor functions of the squared reduced density gradient,
\begin{equation}
p = s^2 = \frac{1}{4(3^{2/3}\pi^{6/3})}\frac{|\nabla n|^2}{n^{8/3}},
\end{equation}
built for the $\alpha = 0$ and $\alpha = 1$ limits as $F^{0}_{\mr{x}}(p)$ and $F^{\mr{1}}_{\mr{x}}(p)$ respectively,
\begin{align}
F^{0}_{\mr{x}}(p) &= 1 + \kappa - \frac{\kappa}{1 + \frac{\mu^{\mr{GE}}p + c}{\kappa}} \\
F^{1}_{\mr{x}}(p) &= 1 + \kappa - \frac{\kappa}{1+ \frac{\mu^{\mr{GE}}p}{\kappa}},
\end{align}
where $\mu^{\mr{GE}} = 10/81$, and $\kappa = 0.504$ and $c = 0.14601$ are parameters determined in Ref. \cite{Sun2013}. $F^{0}_{\mr{x}}$ and $F^{1}_{\mr{x}}$ are independent of $\alpha$, and hence $\beta$, though the $\kappa$ and $c$ parameters should be adjusted accordingly.

The enhancement factor interpolation is a direct function of $\alpha$,
\begin{equation}
\label{eq:falpha}
f(\alpha) = \frac{(1 - \alpha^2)^3}{1 + \alpha^3 + b\alpha^6},
\end{equation}
where $b = 4.0$ is a parameter controlling $\alpha > 1$ behaviour and is optimised simultaneously with $\kappa$ in Ref. \cite{Sun2013} by fitting to atomisation energy and barrier height training sets, AE6 \cite{Lynch2003} and BH6 \cite{Haunschild2012} respectively. Here, we construct a similar interpolation function, exhibiting the same limiting behaviours, in terms of $\beta$. We write,
\begin{align}
\lim_{|\nabla n|\rightarrow 0} f(\alpha) &= \lim_{|\nabla n| \rightarrow 0} \frac{(1-\alpha^2)^3}{1+\alpha^3+b\alpha^6} \\
 &= [-2(-\frac{1}{9}A+\frac{1}{6}B)]^3+ O(\nabla^6) \label{eq:exp}
\end{align}
and this term only contributes to the order of $\nabla^6$ and higher. We can therefore replace $\alpha$ by $2\beta$ in Eq. \ref{eq:falpha} without affecting the second order gradient expansion of the MS2 exchange enhancement factor $F_{\mr{x}}^{\mr{MS2}}$ of Ref. \cite{Sun2013}. We propose,
\begin{equation}
\label{eq:fmbeta}
f_m(\beta) = \frac{\left[1 - (2\beta)^2\right]^3}{1 + (2\beta)^3 + b_m(2\beta)^6},
\end{equation}
where the $m$ subscript indicates a modified term. Firstly, we confirm that Eq. \ref{eq:fmbeta} exhibits the same behaviours as Eq. \ref{eq:falpha} with respect to the equivalent points highlighted in the main text,
\begin{align}
f_m(\beta = 0) &= f(\alpha = 0) = 1, \\
f_m(\beta = 0.5) &= f(\alpha = 1) = 0,
\end{align}
the $\alpha \rightarrow \infty$ limit corresponds to $\beta \rightarrow 1$ giving,
\begin{equation}
f_m(\beta = 1) = \frac{(1 - 4)^3}{1 + 8 + 64b_m}.
\end{equation}
The $f(\alpha)$ function gives,
\begin{align}
f_m(\beta=1) &= \lim_{\alpha \rightarrow \infty} f(\alpha),\\
&= -\frac{1}{b},\\
\end{align}
hence for $f_m(\beta \rightarrow 1)$ to coincide $f(\alpha \rightarrow \infty)$ then the $b_m$ parameter is defined relative to the original $b$ as,
\begin{equation}
b_m = \frac{27b - 9}{64}.
\end{equation}

\section{Atomic Total Energies}

\begin{table}
\caption{The total energies of the neon and argon atoms calculated in the aug-cc-pVQZ basis using a very fine (\textsc{Turbomole} Level 7) integration grid converged to $10^{-8} E_{\mr{h}}$ in the SCF cycle.}
\begin{tabular}{l|cc}
\hline \hline
 & MS2 & MS2$\beta$ \\
\hline
Ne & -128.96625003 & -129.02147785 \\
Ar & -527.53776112 & -527.53776112 \\
\hline \hline
\end{tabular}
\end{table}

\section{Magnetic Fields}

We note that the required modification for kinetic energy density dependent functionals in strong magnetic fields \cite{Bates2012, Furness2015} of substituting the orbital kinetic energy density with the physical kinetic energy density, as derived by Dobson in Ref. \cite{Dobson1992, Dobson1993}, indicator quantity is equally applicable to the $\beta$ quantity as $\alpha$ to construct current sensitive meta-GGA functionals. 

\providecommand{\latin}[1]{#1}
\makeatletter
\providecommand{\doi}
  {\begingroup\let\do\@makeother\dospecials
  \catcode`\{=1 \catcode`\}=2 \doi@aux}
\providecommand{\doi@aux}[1]{\endgroup\texttt{#1}}
\makeatother
\providecommand*\mcitethebibliography{\thebibliography}
\csname @ifundefined\endcsname{endmcitethebibliography}
  {\let\endmcitethebibliography\endthebibliography}{}


\begin{mcitethebibliography}{63}
\providecommand*\natexlab[1]{#1}
\providecommand*\mciteSetBstSublistMode[1]{}
\providecommand*\mciteSetBstMaxWidthForm[2]{}
\providecommand*\mciteBstWouldAddEndPuncttrue
  {\def\EndOfBibitem{\unskip.}}
\providecommand*\mciteBstWouldAddEndPunctfalse
  {\let\EndOfBibitem\relax}
\providecommand*\mciteSetBstMidEndSepPunct[3]{}
\providecommand*\mciteSetBstSublistLabelBeginEnd[3]{}
\providecommand*\EndOfBibitem{}
\mciteSetBstSublistMode{f}
\mciteSetBstMaxWidthForm{subitem}{(\alph{mcitesubitemcount})}
\mciteSetBstSublistLabelBeginEnd
  {\mcitemaxwidthsubitemform\space}
  {\relax}
  {\relax}

\bibitem[Hohenberg and Kohn(1964)Hohenberg, and Kohn]{Hohenberg1964}
Hohenberg,~P.; Kohn,~W. {Inhomogeneous electron gas}. \emph{Physical Review}
  \textbf{1964}, \emph{136}, 864--871\relax
\mciteBstWouldAddEndPuncttrue
\mciteSetBstMidEndSepPunct{\mcitedefaultmidpunct}
{\mcitedefaultendpunct}{\mcitedefaultseppunct}\relax
\EndOfBibitem
\bibitem[Kohn and Sham(1965)Kohn, and Sham]{Kohn1965}
Kohn,~W.; Sham,~L.~J. {Self-consistent equations including exchange and
  correlation effects}. \emph{Physical Review} \textbf{1965}, \emph{140},
  1133--1139\relax
\mciteBstWouldAddEndPuncttrue
\mciteSetBstMidEndSepPunct{\mcitedefaultmidpunct}
{\mcitedefaultendpunct}{\mcitedefaultseppunct}\relax
\EndOfBibitem
\bibitem[Perdew and Schmidt(2001)Perdew, and Schmidt]{Perdew2001}
Perdew,~J.~P.; Schmidt,~K. {Jacob's ladder of density functional approximations
  for the exchange-correlation energy}. \emph{AIP Conference Proceedings}
  \textbf{2001}, \emph{577}, 1--20\relax
\mciteBstWouldAddEndPuncttrue
\mciteSetBstMidEndSepPunct{\mcitedefaultmidpunct}
{\mcitedefaultendpunct}{\mcitedefaultseppunct}\relax
\EndOfBibitem
\bibitem[Ma and Brueckner(1968)Ma, and Brueckner]{Ma1968}
Ma,~S.-K.; Brueckner,~K.~A. {Correlation Energy of an Electron Gas with a
  Slowly Varying High Density}. \emph{Physical Review} \textbf{1968},
  \emph{165}, 18--31\relax
\mciteBstWouldAddEndPuncttrue
\mciteSetBstMidEndSepPunct{\mcitedefaultmidpunct}
{\mcitedefaultendpunct}{\mcitedefaultseppunct}\relax
\EndOfBibitem
\bibitem[{Von Barth} and Hedin(1972){Von Barth}, and Hedin]{vonBarth1972}
{Von Barth},~U.; Hedin,~L. {A local exchange-corremation potentiel for the spin
  polarized case: 1}. \emph{Journal of Physics C: Solid State Physics}
  \textbf{1972}, \emph{1629}, 1629--1642\relax
\mciteBstWouldAddEndPuncttrue
\mciteSetBstMidEndSepPunct{\mcitedefaultmidpunct}
{\mcitedefaultendpunct}{\mcitedefaultseppunct}\relax
\EndOfBibitem
\bibitem[Perdew and Zunger(1981)Perdew, and Zunger]{Perdew1981}
Perdew,~J.~P.; Zunger,~A. {Self-interaction correction to density-functional
  approximations for many-electron systems}. \emph{Physical Review B}
  \textbf{1981}, \emph{23}, 5048--5079\relax
\mciteBstWouldAddEndPuncttrue
\mciteSetBstMidEndSepPunct{\mcitedefaultmidpunct}
{\mcitedefaultendpunct}{\mcitedefaultseppunct}\relax
\EndOfBibitem
\bibitem[Perdew and Wang(1992)Perdew, and Wang]{Perdew1992a}
Perdew,~J.~P.; Wang,~Y. {Accurate and Simple Analytic Representation of the
  Electron-Gas Correlation-Energy}. \emph{Physical Review B} \textbf{1992},
  \emph{45}, 13244--13249\relax
\mciteBstWouldAddEndPuncttrue
\mciteSetBstMidEndSepPunct{\mcitedefaultmidpunct}
{\mcitedefaultendpunct}{\mcitedefaultseppunct}\relax
\EndOfBibitem
\bibitem[Sun \latin{et~al.}(2010)Sun, Perdew, and Seidl]{Sun2010}
Sun,~J.; Perdew,~J.~P.; Seidl,~M. {Correlation energy of the uniform electron
  gas from an interpolation between high- and low-density limits}.
  \emph{Physical Review B} \textbf{2010}, \emph{81}, 085123\relax
\mciteBstWouldAddEndPuncttrue
\mciteSetBstMidEndSepPunct{\mcitedefaultmidpunct}
{\mcitedefaultendpunct}{\mcitedefaultseppunct}\relax
\EndOfBibitem
\bibitem[Becke(1988)]{Becke1988}
Becke,~A.~D. {Density-functional exchange-energy approximation with correct
  asymptotic behavior}. \emph{Physical Review A} \textbf{1988}, \emph{38},
  3098--3100\relax
\mciteBstWouldAddEndPuncttrue
\mciteSetBstMidEndSepPunct{\mcitedefaultmidpunct}
{\mcitedefaultendpunct}{\mcitedefaultseppunct}\relax
\EndOfBibitem
\bibitem[Perdew \latin{et~al.}(1992)Perdew, Chevary, Vosko, Jackson, Pederson,
  Singh, and Fiolhais]{Perdew1992}
Perdew,~J.~P.; Chevary,~J.; Vosko,~S.; Jackson,~K.; Pederson,~M.; Singh,~D.;
  Fiolhais,~C. {Atoms, molecules, solids, and surfaces: Applications of the
  generalized gradient approximation for exchange and correlation}.
  \emph{Physical Review B} \textbf{1992}, \emph{46}, 6671--6687\relax
\mciteBstWouldAddEndPuncttrue
\mciteSetBstMidEndSepPunct{\mcitedefaultmidpunct}
{\mcitedefaultendpunct}{\mcitedefaultseppunct}\relax
\EndOfBibitem
\bibitem[Perdew \latin{et~al.}(1996)Perdew, Burke, and Ernzerhof]{Perdew1996}
Perdew,~J.~P.; Burke,~K.; Ernzerhof,~M. {Generalized Gradient Approximation
  Made Simple.} \emph{Physical Review Letters} \textbf{1996}, \emph{77},
  3865--3868\relax
\mciteBstWouldAddEndPuncttrue
\mciteSetBstMidEndSepPunct{\mcitedefaultmidpunct}
{\mcitedefaultendpunct}{\mcitedefaultseppunct}\relax
\EndOfBibitem
\bibitem[Keal and Tozer(2004)Keal, and Tozer]{Keal2004}
Keal,~T.~W.; Tozer,~D.~J. {A semiempirical generalized gradient approximation
  exchange-correlation functional}. \emph{Journal of Chemical Physics}
  \textbf{2004}, \emph{121}, 5654--5660\relax
\mciteBstWouldAddEndPuncttrue
\mciteSetBstMidEndSepPunct{\mcitedefaultmidpunct}
{\mcitedefaultendpunct}{\mcitedefaultseppunct}\relax
\EndOfBibitem
\bibitem[Armiento and Mattsson(2005)Armiento, and Mattsson]{Armiento2005}
Armiento,~R.; Mattsson,~A.~E. {Functional designed to include surface effects
  in self-consistent density functional theory}. \emph{Physical Review B}
  \textbf{2005}, \emph{72}, 085108\relax
\mciteBstWouldAddEndPuncttrue
\mciteSetBstMidEndSepPunct{\mcitedefaultmidpunct}
{\mcitedefaultendpunct}{\mcitedefaultseppunct}\relax
\EndOfBibitem
\bibitem[Perdew \latin{et~al.}(2008)Perdew, Ruzsinszky, Csonka, Vydrov,
  Scuseria, Constantin, Zhou, and Burke]{Perdew2008a}
Perdew,~J.~P.; Ruzsinszky,~A.; Csonka,~G.~I.; Vydrov,~O.~A.; Scuseria,~G.~E.;
  Constantin,~L.~A.; Zhou,~X.; Burke,~K. {Generalized gradient approximation
  for solids and their surfaces}. \emph{Physical Review Letters} \textbf{2008},
  \emph{100}, 136406\relax
\mciteBstWouldAddEndPuncttrue
\mciteSetBstMidEndSepPunct{\mcitedefaultmidpunct}
{\mcitedefaultendpunct}{\mcitedefaultseppunct}\relax
\EndOfBibitem
\bibitem[Constantin \latin{et~al.}(2011)Constantin, Fabiano, Laricchia, and
  {Della Sala}]{Constantin2011}
Constantin,~L.~A.; Fabiano,~E.; Laricchia,~S.; {Della Sala},~F. {Semiclassical
  neutral atom as a reference system in density functional theory}.
  \emph{Physical Review Letters} \textbf{2011}, \emph{106}, 186406\relax
\mciteBstWouldAddEndPuncttrue
\mciteSetBstMidEndSepPunct{\mcitedefaultmidpunct}
{\mcitedefaultendpunct}{\mcitedefaultseppunct}\relax
\EndOfBibitem
\bibitem[Vela \latin{et~al.}(2012)Vela, Pacheco-kato, G{\'{a}}zquez, Campo, and
  Trickey]{Vela2012}
Vela,~A.; Pacheco-kato,~J.~C.; G{\'{a}}zquez,~J.~L.; Campo,~J.~M.;
  Trickey,~S.~B. {Improved constraint satisfaction in a simple generalized
  gradient approximation exchange functional Improved constraint satisfaction
  in a simple generalized gradient}. \emph{The Journal of Chemical Physics}
  \textbf{2012}, \emph{136}, 144115\relax
\mciteBstWouldAddEndPuncttrue
\mciteSetBstMidEndSepPunct{\mcitedefaultmidpunct}
{\mcitedefaultendpunct}{\mcitedefaultseppunct}\relax
\EndOfBibitem
\bibitem[Becke and Roussel(1989)Becke, and Roussel]{Becke1989}
Becke,~A.~D.; Roussel,~M.~R. {Exchange holes in inhomogeneous systems: A
  coordinate-space model}. \emph{Physical Review A} \textbf{1989}, \emph{39},
  3761--3767\relax
\mciteBstWouldAddEndPuncttrue
\mciteSetBstMidEndSepPunct{\mcitedefaultmidpunct}
{\mcitedefaultendpunct}{\mcitedefaultseppunct}\relax
\EndOfBibitem
\bibitem[{Van Voorhis} and Scuseria(1998){Van Voorhis}, and
  Scuseria]{VanVoorhis1998}
{Van Voorhis},~T.; Scuseria,~G.~E. {A novel form for the exchange-correlation
  energy functional}. \emph{Journal of Chemical Physics} \textbf{1998},
  \emph{109}, 400--410\relax
\mciteBstWouldAddEndPuncttrue
\mciteSetBstMidEndSepPunct{\mcitedefaultmidpunct}
{\mcitedefaultendpunct}{\mcitedefaultseppunct}\relax
\EndOfBibitem
\bibitem[Becke(1998)]{Becke1998}
Becke,~A.~D. {A new inhomogeneity parameter in density-functional theory}.
  \emph{Journal of Chemical Physics} \textbf{1998}, \emph{109},
  2092--2098\relax
\mciteBstWouldAddEndPuncttrue
\mciteSetBstMidEndSepPunct{\mcitedefaultmidpunct}
{\mcitedefaultendpunct}{\mcitedefaultseppunct}\relax
\EndOfBibitem
\bibitem[Perdew \latin{et~al.}(1999)Perdew, Kurth, Zupan, and
  Blaha]{Perdew1999}
Perdew,~J.~P.; Kurth,~S.; Zupan,~A.; Blaha,~P. {Accurate Density Functional
  with Correct Formal Properties: A Step Beyond the Generalized Gradient
  Approximation}. \emph{Physical Review Letters} \textbf{1999}, \emph{82},
  2544--2547\relax
\mciteBstWouldAddEndPuncttrue
\mciteSetBstMidEndSepPunct{\mcitedefaultmidpunct}
{\mcitedefaultendpunct}{\mcitedefaultseppunct}\relax
\EndOfBibitem
\bibitem[Tao \latin{et~al.}(2003)Tao, Perdew, Staroverov, and
  Scuseria]{Tao2003}
Tao,~J.; Perdew,~J.~P.; Staroverov,~V.~N.; Scuseria,~G.~E. {Climbing the
  Density Functional Ladder: Non-Empirical Meta-Generalized Gradient
  Approximation Designed for Molecules and Solids}. \emph{Physical Review
  Letters} \textbf{2003}, \emph{91}, 146401\relax
\mciteBstWouldAddEndPuncttrue
\mciteSetBstMidEndSepPunct{\mcitedefaultmidpunct}
{\mcitedefaultendpunct}{\mcitedefaultseppunct}\relax
\EndOfBibitem
\bibitem[Becke(2003)]{Becke2003}
Becke,~A.~D. {A real-space model of nondynamical correlation}. \emph{Journal of
  Chemical Physics} \textbf{2003}, \emph{119}, 2972--2977\relax
\mciteBstWouldAddEndPuncttrue
\mciteSetBstMidEndSepPunct{\mcitedefaultmidpunct}
{\mcitedefaultendpunct}{\mcitedefaultseppunct}\relax
\EndOfBibitem
\bibitem[Zhao and Truhlar(2006)Zhao, and Truhlar]{Zhao2006}
Zhao,~Y.; Truhlar,~D.~G. {A new local density functional for main-group
  thermochemistry, transition metal bonding, thermochemical kinetics, and
  noncovalent interactions}. \emph{Journal of Chemical Physics} \textbf{2006},
  \emph{125}, 194101\relax
\mciteBstWouldAddEndPuncttrue
\mciteSetBstMidEndSepPunct{\mcitedefaultmidpunct}
{\mcitedefaultendpunct}{\mcitedefaultseppunct}\relax
\EndOfBibitem
\bibitem[Perdew \latin{et~al.}(2009)Perdew, Ruzsinszky, Csonka, Constantin, and
  Sun]{Perdew2009}
Perdew,~J.~P.; Ruzsinszky,~A.; Csonka,~G.~I.; Constantin,~L.~A.; Sun,~J.
  {Workhorse Semilocal Density Functional for Condensed Matter Physics and
  Quantum Chemistry}. \emph{Physical Review Letters} \textbf{2009}, \emph{103},
  026403\relax
\mciteBstWouldAddEndPuncttrue
\mciteSetBstMidEndSepPunct{\mcitedefaultmidpunct}
{\mcitedefaultendpunct}{\mcitedefaultseppunct}\relax
\EndOfBibitem
\bibitem[Peverati and Truhlar(2012)Peverati, and Truhlar]{Peverati2012}
Peverati,~R.; Truhlar,~D.~G. {An improved and broadly accurate local
  approximation to the exchange-correlation density functional: The MN12-L
  functional for electronic structure calculations in chemistry and physics}.
  \emph{Physical Chemistry Chemical Physics} \textbf{2012}, \emph{14},
  13171--13174\relax
\mciteBstWouldAddEndPuncttrue
\mciteSetBstMidEndSepPunct{\mcitedefaultmidpunct}
{\mcitedefaultendpunct}{\mcitedefaultseppunct}\relax
\EndOfBibitem
\bibitem[Yu \latin{et~al.}(2016)Yu, He, and Truhlar]{Yu2016a}
Yu,~H.~S.; He,~X.; Truhlar,~D.~G. {MN15-L: A New Local Exchange-Correlation
  Functional for Kohn-Sham Density Functional Theory with Broad Accuracy for
  Atoms, Molecules, and Solids}. \emph{Journal of Chemical Theory and
  Computation} \textbf{2016}, \emph{12}, 1280--1293\relax
\mciteBstWouldAddEndPuncttrue
\mciteSetBstMidEndSepPunct{\mcitedefaultmidpunct}
{\mcitedefaultendpunct}{\mcitedefaultseppunct}\relax
\EndOfBibitem
\bibitem[Sun \latin{et~al.}(2012)Sun, Xiao, and Ruzsinszky]{Sun2012}
Sun,~J.; Xiao,~B.; Ruzsinszky,~A. {Communication : Effect of the
  orbital-overlap dependence in the meta generalized gradient approximation}.
  \emph{Journal of Chemical Physics} \textbf{2012}, \emph{137}, 051101\relax
\mciteBstWouldAddEndPuncttrue
\mciteSetBstMidEndSepPunct{\mcitedefaultmidpunct}
{\mcitedefaultendpunct}{\mcitedefaultseppunct}\relax
\EndOfBibitem
\bibitem[{Del Campo} \latin{et~al.}(2012){Del Campo}, G{\'{a}}zquez, Trickey,
  and Vela]{DelCampo2012}
{Del Campo},~J.~M.; G{\'{a}}zquez,~J.~L.; Trickey,~S.~B.; Vela,~A. {A new
  meta-GGA exchange functional based on an improved constraint-based GGA}.
  \emph{Chemical Physics Letters} \textbf{2012}, \emph{543}, 179--183\relax
\mciteBstWouldAddEndPuncttrue
\mciteSetBstMidEndSepPunct{\mcitedefaultmidpunct}
{\mcitedefaultendpunct}{\mcitedefaultseppunct}\relax
\EndOfBibitem
\bibitem[Mardirossian and Head-Gordon(2015)Mardirossian, and
  Head-Gordon]{Mardirossian2015}
Mardirossian,~N.; Head-Gordon,~M. {Mapping the genome of meta-generalized
  gradient approximation density functionals: The search for B97M-V}.
  \emph{Journal of Chemical Physics} \textbf{2015}, \emph{142}, 074111\relax
\mciteBstWouldAddEndPuncttrue
\mciteSetBstMidEndSepPunct{\mcitedefaultmidpunct}
{\mcitedefaultendpunct}{\mcitedefaultseppunct}\relax
\EndOfBibitem
\bibitem[Sun \latin{et~al.}(2015)Sun, Ruzsinszky, and Perdew]{Sun2015}
Sun,~J.; Ruzsinszky,~A.; Perdew,~J.~P. {Strongly Constrained and Appropriately
  Normed Semilocal Density Functional}. \emph{Physical Review Letters}
  \textbf{2015}, \emph{115}, 036402\relax
\mciteBstWouldAddEndPuncttrue
\mciteSetBstMidEndSepPunct{\mcitedefaultmidpunct}
{\mcitedefaultendpunct}{\mcitedefaultseppunct}\relax
\EndOfBibitem
\bibitem[Perdew \latin{et~al.}(2006)Perdew, Constantin, Sagvolden, and
  Burke]{Perdew2006}
Perdew,~J.~P.; Constantin,~L.~A.; Sagvolden,~E.; Burke,~K. {Relevance of the
  slowly varying electron gas to atoms, molecules, and solids}. \emph{Physical
  Review Letters} \textbf{2006}, \emph{97}, 1--4\relax
\mciteBstWouldAddEndPuncttrue
\mciteSetBstMidEndSepPunct{\mcitedefaultmidpunct}
{\mcitedefaultendpunct}{\mcitedefaultseppunct}\relax
\EndOfBibitem
\bibitem[Becke(1983)]{Becke1983}
Becke,~A.~D. {Hartree-Fock exchange energy of an inhomogeneous electron gas}.
  \emph{International Journal of Quantum Chemistry} \textbf{1983}, \emph{23},
  1915--1922\relax
\mciteBstWouldAddEndPuncttrue
\mciteSetBstMidEndSepPunct{\mcitedefaultmidpunct}
{\mcitedefaultendpunct}{\mcitedefaultseppunct}\relax
\EndOfBibitem
\bibitem[Wang \latin{et~al.}(2017)Wang, Jin, Yu, Truhlar, and He]{Wang2017}
Wang,~Y.; Jin,~X.; Yu,~H.~S.; Truhlar,~D.~G.; He,~X. {Revised M06-L functional
  for improved accuracy on chemical reaction barrier heights, noncovalent
  interactions, and solid-state physics}. \emph{Proceedings of the National
  Academy of Sciences} \textbf{2017}, \emph{114}, 8487--8492\relax
\mciteBstWouldAddEndPuncttrue
\mciteSetBstMidEndSepPunct{\mcitedefaultmidpunct}
{\mcitedefaultendpunct}{\mcitedefaultseppunct}\relax
\EndOfBibitem
\bibitem[Tao and Mo(2016)Tao, and Mo]{Tao2016}
Tao,~J.; Mo,~Y. {Accurate Semilocal Density Functional for Condensed-Matter
  Physics and Quantum Chemistry}. \emph{Physical Review Letters} \textbf{2016},
  \emph{117}, 073001\relax
\mciteBstWouldAddEndPuncttrue
\mciteSetBstMidEndSepPunct{\mcitedefaultmidpunct}
{\mcitedefaultendpunct}{\mcitedefaultseppunct}\relax
\EndOfBibitem
\bibitem[Sun \latin{et~al.}(2013)Sun, Xiao, Fang, Haunschild, Hao, Ruzsinszky,
  Csonka, Scuseria, and Perdew]{Sun2013a}
Sun,~J.; Xiao,~B.; Fang,~Y.; Haunschild,~R.; Hao,~P.; Ruzsinszky,~A.;
  Csonka,~G.~I.; Scuseria,~G.~E.; Perdew,~J.~P. {Density functionals that
  recognize covalent, metallic, and weak bonds}. \emph{Physical Review Letters}
  \textbf{2013}, \emph{111}, 106401\relax
\mciteBstWouldAddEndPuncttrue
\mciteSetBstMidEndSepPunct{\mcitedefaultmidpunct}
{\mcitedefaultendpunct}{\mcitedefaultseppunct}\relax
\EndOfBibitem
\bibitem[Svendsen and von Barth(1996)Svendsen, and von Barth]{Svendsen1996}
Svendsen,~P.; von Barth,~U. {Gradient expansion of the exchange energy from
  second-order density response theory}. \emph{Physical Review B}
  \textbf{1996}, \emph{54}, 17402--17413\relax
\mciteBstWouldAddEndPuncttrue
\mciteSetBstMidEndSepPunct{\mcitedefaultmidpunct}
{\mcitedefaultendpunct}{\mcitedefaultseppunct}\relax
\EndOfBibitem
\bibitem[Becke and Edgecombe(1990)Becke, and Edgecombe]{Becke1990}
Becke,~A.~D.; Edgecombe,~K.~E. {A simple measure of electron localization in
  atomic and molecular systems}. \emph{The Journal of Chemical Physics}
  \textbf{1990}, \emph{92}, 5397--5403\relax
\mciteBstWouldAddEndPuncttrue
\mciteSetBstMidEndSepPunct{\mcitedefaultmidpunct}
{\mcitedefaultendpunct}{\mcitedefaultseppunct}\relax
\EndOfBibitem
\bibitem[Silvi and Savin(1994)Silvi, and Savin]{Silvi1994}
Silvi,~B.; Savin,~A. {Classification of Chemical-Bonds Based on Topological
  Analysis of Electron Localization Functions}. \emph{Nature} \textbf{1994},
  \emph{371}, 683--686\relax
\mciteBstWouldAddEndPuncttrue
\mciteSetBstMidEndSepPunct{\mcitedefaultmidpunct}
{\mcitedefaultendpunct}{\mcitedefaultseppunct}\relax
\EndOfBibitem
\bibitem[Savin \latin{et~al.}(1996)Savin, Silvi, and Colonna]{Savin1996}
Savin,~A.; Silvi,~B.; Colonna,~F. {Topological analysis of the electron
  localization function applied to delocalized bonds}. \emph{Canadian Journal
  of Chemistry} \textbf{1996}, \emph{74}, 1088--1096\relax
\mciteBstWouldAddEndPuncttrue
\mciteSetBstMidEndSepPunct{\mcitedefaultmidpunct}
{\mcitedefaultendpunct}{\mcitedefaultseppunct}\relax
\EndOfBibitem
\bibitem[Savin \latin{et~al.}(1997)Savin, Nesper, Wengert, and
  F{\"{a}}ssler]{Savin1997}
Savin,~A.; Nesper,~R.; Wengert,~S.; F{\"{a}}ssler,~T.~F. {ELF: The Electron
  Localization Function}. \emph{Angewandte Chemie International Edition in
  English} \textbf{1997}, \emph{36}, 1808--1832\relax
\mciteBstWouldAddEndPuncttrue
\mciteSetBstMidEndSepPunct{\mcitedefaultmidpunct}
{\mcitedefaultendpunct}{\mcitedefaultseppunct}\relax
\EndOfBibitem
\bibitem[Noury \latin{et~al.}(1998)Noury, Colonna, Savin, and Silvi]{Noury1998}
Noury,~S.; Colonna,~F.; Savin,~A.; Silvi,~B. {Analysis of the delocalization in
  the topological theory of chemical bond}. \emph{Journal of Molecular
  Structure} \textbf{1998}, \emph{450}, 59--68\relax
\mciteBstWouldAddEndPuncttrue
\mciteSetBstMidEndSepPunct{\mcitedefaultmidpunct}
{\mcitedefaultendpunct}{\mcitedefaultseppunct}\relax
\EndOfBibitem
\bibitem[Paul \latin{et~al.}(2017)Paul, Sun, Perdew, and Waghmare]{Paul2017}
Paul,~A.; Sun,~J.; Perdew,~J.~P.; Waghmare,~U.~V. {Accuracy of first-principles
  interatomic interactions and predictions of ferroelectric phase transitions
  in perovskite oxides: Energy functional and effective Hamiltonian}.
  \emph{Physical Review B} \textbf{2017}, \emph{95}, 054111\relax
\mciteBstWouldAddEndPuncttrue
\mciteSetBstMidEndSepPunct{\mcitedefaultmidpunct}
{\mcitedefaultendpunct}{\mcitedefaultseppunct}\relax
\EndOfBibitem
\bibitem[Zhang \latin{et~al.}(2017)Zhang, Sun, Perdew, and Wu]{Zhang2017a}
Zhang,~Y.; Sun,~J.; Perdew,~J.~P.; Wu,~X. {Comparative first-principles studies
  of prototypical ferroelectric materials by LDA, GGA, and SCAN meta-GGA}.
  \emph{Physical Review B} \textbf{2017}, \emph{96}, 035143\relax
\mciteBstWouldAddEndPuncttrue
\mciteSetBstMidEndSepPunct{\mcitedefaultmidpunct}
{\mcitedefaultendpunct}{\mcitedefaultseppunct}\relax
\EndOfBibitem
\bibitem[Zhang \latin{et~al.}(2018)Zhang, Kitchaev, Yang, Chen, Dacek,
  Sarmiento-P{\'{e}}rez, Marques, Peng, Ceder, Perdew, and Sun]{Zhang2018}
Zhang,~Y.; Kitchaev,~D.~A.; Yang,~J.; Chen,~T.; Dacek,~S.~T.;
  Sarmiento-P{\'{e}}rez,~R.~A.; Marques,~M. A.~L.; Peng,~H.; Ceder,~G.;
  Perdew,~J.~P. \latin{et~al.}  {Efficient first-principles prediction of solid
  stability: Towards chemical accuracy}. \emph{npj Computational Materials}
  \textbf{2018}, \emph{4}\relax
\mciteBstWouldAddEndPuncttrue
\mciteSetBstMidEndSepPunct{\mcitedefaultmidpunct}
{\mcitedefaultendpunct}{\mcitedefaultseppunct}\relax
\EndOfBibitem
\bibitem[Sun \latin{et~al.}(2016)Sun, Remsing, Zhang, Sun, Ruzsinszky, Peng,
  Yang, Paul, Waghmare, Wu, Klein, and Perdew]{Sun2016}
Sun,~J.; Remsing,~R.~C.; Zhang,~Y.; Sun,~Z.; Ruzsinszky,~A.; Peng,~H.;
  Yang,~Z.; Paul,~A.; Waghmare,~U.; Wu,~X. \latin{et~al.}  {Accurate
  first-principles structures and energies of diversely bonded systems from an
  efficient density functional}. \emph{Nature Chemistry} \textbf{2016},
  \emph{8}, 831--836\relax
\mciteBstWouldAddEndPuncttrue
\mciteSetBstMidEndSepPunct{\mcitedefaultmidpunct}
{\mcitedefaultendpunct}{\mcitedefaultseppunct}\relax
\EndOfBibitem
\bibitem[Furness \latin{et~al.}(2018)Furness, Zhang, Lane, Buda, Barbiellini,
  Markiewicz, Bansil, and Sun]{Furness2018}
Furness,~J.~W.; Zhang,~Y.; Lane,~C.; Buda,~I.~G.; Barbiellini,~B.;
  Markiewicz,~R.~S.; Bansil,~A.; Sun,~J. {An accurate first-principles
  treatment of doping-dependent electronic structure of high-temperature
  cuprate superconductors}. \emph{Communications Physics} \textbf{2018},
  \emph{1}, 1--6\relax
\mciteBstWouldAddEndPuncttrue
\mciteSetBstMidEndSepPunct{\mcitedefaultmidpunct}
{\mcitedefaultendpunct}{\mcitedefaultseppunct}\relax
\EndOfBibitem
\bibitem[Yang \latin{et~al.}(2016)Yang, Peng, Sun, and Perdew]{Yang2016}
Yang,~Z.-h.; Peng,~H.; Sun,~J.; Perdew,~J.~P. {More realistic band gaps from
  meta-generalized gradient approximations: Only in a generalized Kohn-Sham
  scheme}. \emph{Physical Review B} \textbf{2016}, \emph{93}, 205205\relax
\mciteBstWouldAddEndPuncttrue
\mciteSetBstMidEndSepPunct{\mcitedefaultmidpunct}
{\mcitedefaultendpunct}{\mcitedefaultseppunct}\relax
\EndOfBibitem
\bibitem[Johnson \latin{et~al.}(2009)Johnson, Becke, Sherrill, and
  Dilabio]{Johnson2009}
Johnson,~E.~R.; Becke,~A.~D.; Sherrill,~C.~D.; Dilabio,~G.~A. {Oscillations in
  meta-generalized-gradient approximation potential energy surfaces for
  dispersion-bound complexes}. \emph{The Journal of Chemical Physics}
  \textbf{2009}, \emph{131}, 034111\relax
\mciteBstWouldAddEndPuncttrue
\mciteSetBstMidEndSepPunct{\mcitedefaultmidpunct}
{\mcitedefaultendpunct}{\mcitedefaultseppunct}\relax
\EndOfBibitem
\bibitem[Mardirossian and Head-Gordon(2017)Mardirossian, and
  Head-Gordon]{Mardirossian2017a}
Mardirossian,~N.; Head-Gordon,~M. {Thirty years of density functional theory in
  computational chemistry: An overview and extensive assessment of 200 density
  functionals}. \emph{Molecular Physics} \textbf{2017}, \emph{115},
  2315--2372\relax
\mciteBstWouldAddEndPuncttrue
\mciteSetBstMidEndSepPunct{\mcitedefaultmidpunct}
{\mcitedefaultendpunct}{\mcitedefaultseppunct}\relax
\EndOfBibitem
\bibitem[Perdew \latin{et~al.}(2014)Perdew, Ruzsinszky, Sun, and
  Burke]{Perdew2014}
Perdew,~J.~P.; Ruzsinszky,~A.; Sun,~J.; Burke,~K. {Gedanken densities and exact
  constraints in density functional theory}. \emph{The Journal of Chemical
  Physics} \textbf{2014}, \emph{140}, 18A533\relax
\mciteBstWouldAddEndPuncttrue
\mciteSetBstMidEndSepPunct{\mcitedefaultmidpunct}
{\mcitedefaultendpunct}{\mcitedefaultseppunct}\relax
\EndOfBibitem
\bibitem[Sun \latin{et~al.}(2015)Sun, Perdew, and Ruzsinszky]{Sun2015a}
Sun,~J.; Perdew,~J.~P.; Ruzsinszky,~A. {Semilocal density functional obeying a
  strongly tightened bound for exchange}. \emph{Proceedings of the National
  Academy of Sciences} \textbf{2015}, \emph{112}, 685--689\relax
\mciteBstWouldAddEndPuncttrue
\mciteSetBstMidEndSepPunct{\mcitedefaultmidpunct}
{\mcitedefaultendpunct}{\mcitedefaultseppunct}\relax
\EndOfBibitem
\bibitem[Sun \latin{et~al.}(2013)Sun, Haunschild, Xiao, Bulik, Scuseria, and
  Perdew]{Sun2013}
Sun,~J.; Haunschild,~R.; Xiao,~B.; Bulik,~I.~W.; Scuseria,~G.~E.; Perdew,~J.~P.
  {Semilocal and hybrid meta-generalized gradient approximations based on the
  understanding of the kinetic-energy-density dependence Semilocal and hybrid
  meta-generalized gradient approximations based on the understanding of the
  kinetic-energy-density depend}. \emph{Journal of Chemical Physics}
  \textbf{2013}, \emph{138}, 044113\relax
\mciteBstWouldAddEndPuncttrue
\mciteSetBstMidEndSepPunct{\mcitedefaultmidpunct}
{\mcitedefaultendpunct}{\mcitedefaultseppunct}\relax
\EndOfBibitem
\bibitem[TUR()]{TURBOMOLE}
{TURBOMOLE V7.2 2018, a development of University of Karlsruhe and
  Forschungszentrum Karlsruhe GmbH, 1989-2007, TURBOMOLE GmbH, since 2007;
  available from http://www.turbomole.com.}\relax
\mciteBstWouldAddEndPunctfalse
\mciteSetBstMidEndSepPunct{\mcitedefaultmidpunct}
{}{\mcitedefaultseppunct}\relax
\EndOfBibitem
\bibitem[Ekstr{\"{o}}m \latin{et~al.}(2010)Ekstr{\"{o}}m, Visscher, Bast, and
  Thorvaldsen]{Ekstrom2010}
Ekstr{\"{o}}m,~U.; Visscher,~L.; Bast,~R.; Thorvaldsen,~A.~J. {Arbitrary-Order
  Density Functional Response Theory}. \emph{Journal of Chemical Theory and
  Computation} \textbf{2010}, \emph{6}, 1971--1980\relax
\mciteBstWouldAddEndPuncttrue
\mciteSetBstMidEndSepPunct{\mcitedefaultmidpunct}
{\mcitedefaultendpunct}{\mcitedefaultseppunct}\relax
\EndOfBibitem
\bibitem[Dunning(1989)]{Dunning1989}
Dunning,~T.~H. {Gaussian basis sets for use in correlated molecular
  calculations. I. The atoms boron through neon and hydrogen}. \emph{The
  Journal of Chemical Physics} \textbf{1989}, \emph{90}, 1007--1023\relax
\mciteBstWouldAddEndPuncttrue
\mciteSetBstMidEndSepPunct{\mcitedefaultmidpunct}
{\mcitedefaultendpunct}{\mcitedefaultseppunct}\relax
\EndOfBibitem
\bibitem[Neumann \latin{et~al.}(1996)Neumann, Nobes, and Handy]{Neumann1996}
Neumann,~R.; Nobes,~R.~H.; Handy,~N.~C. {Exchange functionals and potentials}.
  \emph{Molecular Physics} \textbf{1996}, \emph{87}, 1--36\relax
\mciteBstWouldAddEndPuncttrue
\mciteSetBstMidEndSepPunct{\mcitedefaultmidpunct}
{\mcitedefaultendpunct}{\mcitedefaultseppunct}\relax
\EndOfBibitem
\bibitem[Seidl \latin{et~al.}(1996)Seidl, G{\"{o}}rling, Vogl, Majewski, and
  Levy]{Seidl1996}
Seidl,~A.; G{\"{o}}rling,~A.; Vogl,~P.; Majewski,~J.; Levy,~M. {Generalized
  Kohn-Sham schemes and the band-gap problem}. \emph{Physical Review B}
  \textbf{1996}, \emph{53}, 3764--3774\relax
\mciteBstWouldAddEndPuncttrue
\mciteSetBstMidEndSepPunct{\mcitedefaultmidpunct}
{\mcitedefaultendpunct}{\mcitedefaultseppunct}\relax
\EndOfBibitem
\bibitem[Clark \latin{et~al.}(1983)Clark, Chandrasekhar, Spitznagel, and
  Schleyer]{Clark1983}
Clark,~T.; Chandrasekhar,~J.; Spitznagel,~G.~W.; Schleyer,~P. V.~R. {Efficient
  diffuse function-augmented basis sets for anion calculations. III. The 3-21+G
  basis set for first-row elements, Li-F}. \emph{Journal of Computational
  Chemistry} \textbf{1983}, \emph{4}, 294--301\relax
\mciteBstWouldAddEndPuncttrue
\mciteSetBstMidEndSepPunct{\mcitedefaultmidpunct}
{\mcitedefaultendpunct}{\mcitedefaultseppunct}\relax
\EndOfBibitem
\bibitem[Frisch \latin{et~al.}(1984)Frisch, Pople, and Binkley]{Frisch1984a}
Frisch,~M.~J.; Pople,~J.~A.; Binkley,~J.~S. {Self-consistent molecular orbital
  methods 25. Supplementary functions for Gaussian basis sets}. \emph{The
  Journal of Chemical Physics} \textbf{1984}, \emph{80}, 3265--3269\relax
\mciteBstWouldAddEndPuncttrue
\mciteSetBstMidEndSepPunct{\mcitedefaultmidpunct}
{\mcitedefaultendpunct}{\mcitedefaultseppunct}\relax
\EndOfBibitem
\bibitem[Lynch and Truhlar(2003)Lynch, and Truhlar]{Lynch2003}
Lynch,~B.~J.; Truhlar,~D.~G. {Small Representative Benchmarks for
  Thermochemical Calculations}. \emph{Journal of Physical Chemistry A}
  \textbf{2003}, \emph{107}, 8996--8999\relax
\mciteBstWouldAddEndPuncttrue
\mciteSetBstMidEndSepPunct{\mcitedefaultmidpunct}
{\mcitedefaultendpunct}{\mcitedefaultseppunct}\relax
\EndOfBibitem
\bibitem[Haunschild and Klopper(2012)Haunschild, and Klopper]{Haunschild2012}
Haunschild,~R.; Klopper,~W. {Theoretical reference values for the AE6 and BH6
  test sets from explicitly correlated coupled-cluster theory}.
  \emph{Theoretical Chemistry Accounts} \textbf{2012}, \emph{131}, 1112\relax
\mciteBstWouldAddEndPuncttrue
\mciteSetBstMidEndSepPunct{\mcitedefaultmidpunct}
{\mcitedefaultendpunct}{\mcitedefaultseppunct}\relax
\EndOfBibitem
\bibitem[Patkowski \latin{et~al.}(2005)Patkowski, Murdachaew, Fou, and
  Szalewicz]{Patkowski2005}
Patkowski,~K.; Murdachaew,~G.; Fou,~C.~M.; Szalewicz,~K. {Accurate ab initio
  potential for argon dimer including highly repulsive region}. \emph{Molecular
  Physics} \textbf{2005}, \emph{103}, 2031--2045\relax
\mciteBstWouldAddEndPuncttrue
\mciteSetBstMidEndSepPunct{\mcitedefaultmidpunct}
{\mcitedefaultendpunct}{\mcitedefaultseppunct}\relax
\EndOfBibitem
\end{mcitethebibliography}

\begin{mcitethebibliography}{11}
\providecommand*\natexlab[1]{#1}
\providecommand*\mciteSetBstSublistMode[1]{}
\providecommand*\mciteSetBstMaxWidthForm[2]{}
\providecommand*\mciteBstWouldAddEndPuncttrue
  {\def\EndOfBibitem{\unskip.}}
\providecommand*\mciteBstWouldAddEndPunctfalse
  {\let\EndOfBibitem\relax}
\providecommand*\mciteSetBstMidEndSepPunct[3]{}
\providecommand*\mciteSetBstSublistLabelBeginEnd[3]{}
\providecommand*\EndOfBibitem{}
\mciteSetBstSublistMode{f}
\mciteSetBstMaxWidthForm{subitem}{(\alph{mcitesubitemcount})}
\mciteSetBstSublistLabelBeginEnd
  {\mcitemaxwidthsubitemform\space}
  {\relax}
  {\relax}

\bibitem[Jure{\v{c}}ka \latin{et~al.}(2006)Jure{\v{c}}ka, {\v{S}}poner,
  {\v{C}}ern{\'{y}}, and Hobza]{Jurecka2006}
Jure{\v{c}}ka,~P.; {\v{S}}poner,~J.; {\v{C}}ern{\'{y}},~J.; Hobza,~P.
  {Benchmark database of accurate (MP2 and CCSD(T) complete basis set limit)
  interaction energies of small model complexes, DNA base pairs, and amino acid
  pairs}. \emph{Physical Chemistry Chemical Physics} \textbf{2006}, \emph{8},
  1985--1993\relax
\mciteBstWouldAddEndPuncttrue
\mciteSetBstMidEndSepPunct{\mcitedefaultmidpunct}
{\mcitedefaultendpunct}{\mcitedefaultseppunct}\relax
\EndOfBibitem
\bibitem[Becke(1993)]{Becke1993}
Becke,~A.~D. {A new mixing of Hartree–Fock and local density-functional
  theories}. \emph{The Journal of Chemical Physics} \textbf{1993}, \emph{98},
  1372\relax
\mciteBstWouldAddEndPuncttrue
\mciteSetBstMidEndSepPunct{\mcitedefaultmidpunct}
{\mcitedefaultendpunct}{\mcitedefaultseppunct}\relax
\EndOfBibitem
\bibitem[Bates and Furche(2012)Bates, and Furche]{Bates2012}
Bates,~J.~E.; Furche,~F. {Harnessing the meta-generalized gradient
  approximation for time-dependent density functional theory}. \emph{Journal of
  Chemical Physics} \textbf{2012}, \emph{137}, 164105\relax
\mciteBstWouldAddEndPuncttrue
\mciteSetBstMidEndSepPunct{\mcitedefaultmidpunct}
{\mcitedefaultendpunct}{\mcitedefaultseppunct}\relax
\EndOfBibitem
\bibitem[Furness \latin{et~al.}(2015)Furness, Verbeke, Tellgren, Stopkowicz,
  Ekstr{\"{o}}m, Helgaker, and Teale]{Furness2015}
Furness,~J.~W.; Verbeke,~J.; Tellgren,~E.~I.; Stopkowicz,~S.;
  Ekstr{\"{o}}m,~U.; Helgaker,~T.; Teale,~A.~M. {Current Density Functional
  Theory Using Meta-Generalized Gradient Exchange-Correlation Functionals}.
  \emph{Journal of Chemical Theory and Computation} \textbf{2015}, \emph{11},
  4169--4181\relax
\mciteBstWouldAddEndPuncttrue
\mciteSetBstMidEndSepPunct{\mcitedefaultmidpunct}
{\mcitedefaultendpunct}{\mcitedefaultseppunct}\relax
\EndOfBibitem
\bibitem[Dobson(1992)]{Dobson1992}
Dobson,~J.~F. {Spin-density functionals for the electron correlation energy
  with automatic freedom from orbital self-interaction}. \emph{Journal of
  Physics: Condensed Matter} \textbf{1992}, \emph{4}, 7877--7890\relax
\mciteBstWouldAddEndPuncttrue
\mciteSetBstMidEndSepPunct{\mcitedefaultmidpunct}
{\mcitedefaultendpunct}{\mcitedefaultseppunct}\relax
\EndOfBibitem
\bibitem[Dobson(1993)]{Dobson1993}
Dobson,~J.~F. {Alternative expressions for the Fermi hole curvature}. \emph{The
  Journal of Chemical Physics} \textbf{1993}, \emph{98}, 8870--8872\relax
\mciteBstWouldAddEndPuncttrue
\mciteSetBstMidEndSepPunct{\mcitedefaultmidpunct}
{\mcitedefaultendpunct}{\mcitedefaultseppunct}\relax
\EndOfBibitem
\end{mcitethebibliography}
\end{document}